\documentclass[preprint,aps,showpacs]{revtex4}
\usepackage{graphicx}
\usepackage{bm}
\usepackage{amsmath}
\newcommand{\nn}{\nonumber}
\begin{document}

\title{Analytical and Experimental Study of X-ray Absorption Coefficients of Material by
Abel's Inversion\footnote{This study was performed for DAHRT
project by the author at LANL (LA-UR-91-1966) and was supported by
DOE. }}

\author{S. J. Han}

\affiliation{P.O. Box 4684, Los Alamos, NM 87544-4684 \\
email:sjhan@cybermesa.com}

\begin{abstract}

The hard x-ray ($\gamma$-ray) absorption by cylindrically
symmetric $U^{238}$ test objects is studied by means of
$\gamma$-ray transmission measurements. To make a precise
comparison between the theoretically modelled values and the
absorption coefficients calculated from the experimental data, we
have developed a highly accurate numerical code based on a new
solution of Abel's integral equation. It is shown that progressive
filtering, surface reflections by Compton scattering, and the
enhanced backscattering due to  impurities can explain much of the
observed discrepancy. We also discuss optimal experimental
conditions with regard to the feasibility of quantitative
radiography for $\gamma$-ray diagnostics.
\end{abstract}

\maketitle

\section{\label{sec:level1} Introduction}

The calculation of precise radial density profiles for a localized
radiation source ({\it i.e.,} soft x-ray emission by hot plasmas)
has been a long standing problem in both plasma physics and
astrophysics
\cite{Horman35,Maecker54,Friedrich59,Bockasten61,Bracewell56}. For
a sufficiently tenuous plasma in which radiation emission
processes dominate over absorption processes, the symmetric
density profile of the plasma can be obtained from the projected
radiation intensity by Abel's inversion.

Another example of the application of Abel's inversion is the
determination of density profiles of an imploding spherical
targets driven by laser-beams \cite{Chang81}. These two seemingly
different diagnostic methods are mathematically identical and rely
on the solutions of Abel's integral equation
\cite{Horman35,Maecker54,Friedrich59,Bockasten61,Bracewell56}.

Before proceeding further, it is important to point out that there
have been extensive studies on the absorption of a soft x-ray by
the Abel's integral equation for medical applications by Cormack
\cite{Cormack63} who apparently was not aware of previous work on
the x-ray diagnostics employed in Astrophysics
\cite{Horman35,Maecker54,Friedrich59,Bockasten61,Bracewell56}.
Despite numerous studies over many years \cite{Peebles84}, there
still exists considerable debate over the importance of a
numerical method for solving the integral equation. The numerical
algorithm by Barr \cite{Barr62} has attracted considerable
interest, particularly with the observation of a diffuse
background due to Compton scattering.

In all x-ray diagnostics, the x-rays are normally produced by a
high-energy electron accelerator and depend on the bremsstrahlung
process in a high Z-material. In particular, the PHERMEX electron
accelerator at Los Alamos that is a 50 MHz, 30 MeV standing-wave
device which operates at 500 A with 600 kV injector. It has been
relatively a simple task to focus the high current electron beam
by injecting a neutral plasma by means of self-focusing by a self-
pinching mechanism \cite{Han1} and thus creates a point-like high
energy x-ray (or $\gamma$-ray) source. The x-rays are produced by
the scattering of the high-energy electron beam at a target
material (usually high Z-material such as tungsten). The electrons
lose their energy by two basic process: collision and radiation.
The collision process leaves the atoms in the material either
excited states or the ionized states. The ejected electrons have a
small energy that is deposited locally. On the other hand the
x-rays (radiation energy) generated by bremsstrahlung process is
uniformly distributed among secondary photons of all energies up
to the energy of the primary electron energy itself.

\begin{figure}[tbp]
\includegraphics[bb=0in 0in 6in 4in,totalheight=4in,height=4in,width=6in,keepaspectratio=true]{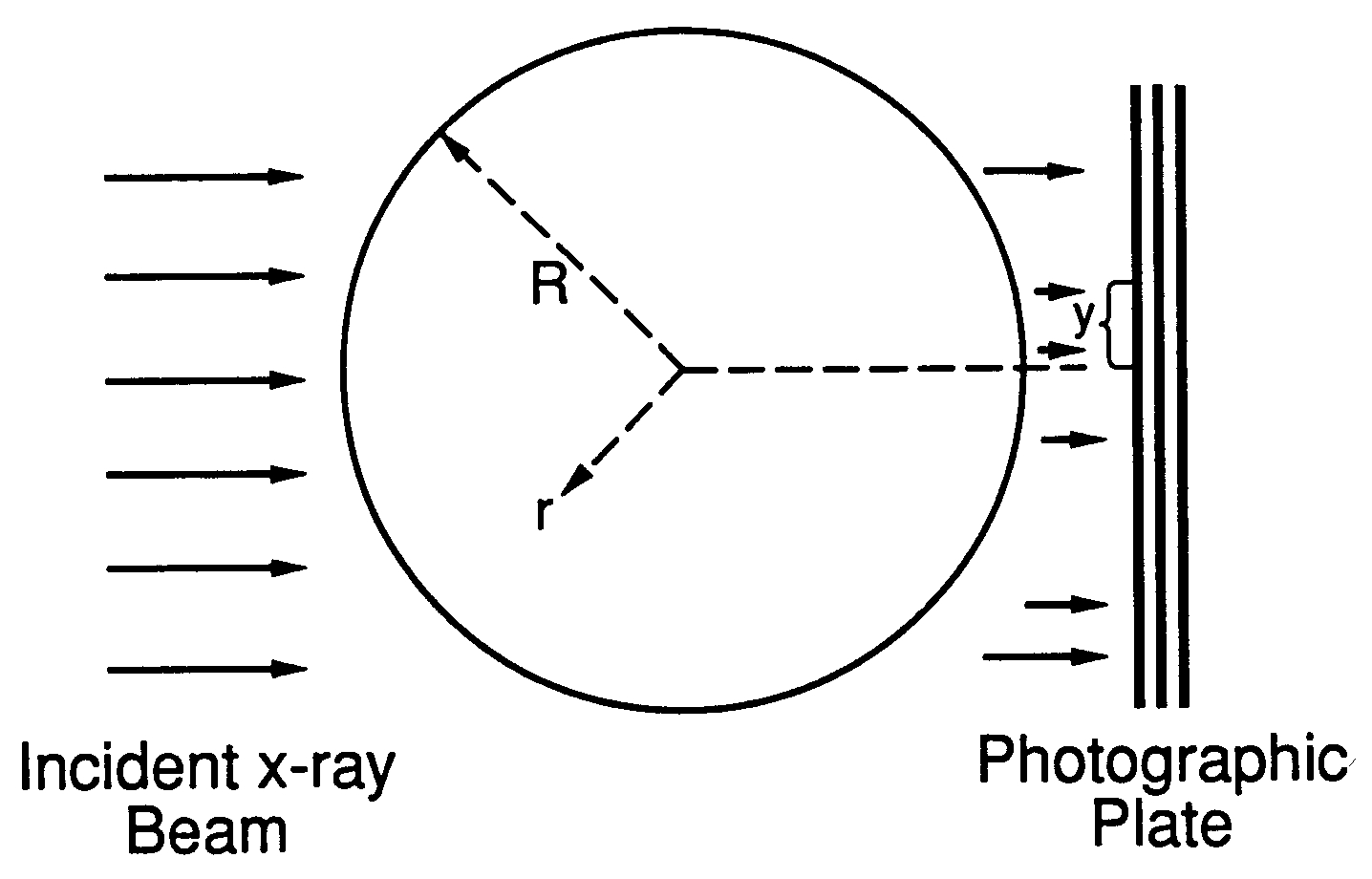}
\vspace{0.0truecm} \caption{A schematic picture of x-ray
transmission experiments. The incident x-ray beam is approximated
to be parallel uniform by the assumption that the source is very
far from the test objects.} \label{fig:fig1}
\end{figure}

At the energy of PHERMEX electron beam, the bremsstrahlung process
dominates over the collision process. For an optimal radiography,
it is essential to have a pencil beam to create a point source of
the x-rays at a distance far from the test object. This provides
approximately a parallel x-ray beam as shown in Figure
\ref{fig:fig1}. Moreover, the spot size of a point source
determines the spatial resolution in an X-ray film.

This optimal condition allows one to study a density profile of a
test object by means of Abel's integral equation
\cite{Comments2}. In all previous numerical calculations based on
the Volterra solution of Abel integral equation \cite{Horman35},
expressing the data as a well-behaved function was an essential
step to avoid the noise amplification. When the data, which
contain small random errors due to a photon cascade and large
fluctuations by the enhanced backscattering at the impurities,
are smoothed by polynomial fits, the calculations tend to obscure
the basic processes. Moreover, the smoothing leads to a
hybridization of random errors and fluctuations such that the
computed absorption coefficients do not provide any meaningful
information about absorption coefficients. Worse yet, the results
are often misleading.

Here we report a study, by flash x-ray photography, of the
absorption coefficients of metallic cylinders of high Z-material
($U^{238}$) with a high concentration of impurities. First we
discuss a new numerical solution to Abel's integral equation and
its application to a cylindrically symmetric object using
experimental data. We then discuss the results, their relevance,
and connection to other work. Last we discuss optimal experimental
conditions with regard to feasibility of x-ray radiography as a
quantitative tool for x-ray diagnostics.

X-ray absorptions at the surface of condensed matter in photon
energies below 40~KeV has been extensively studied. Experimental
measurements of the absorption coefficients are in good agreement
with detailed theoretical analysis \cite{Prins88}. However, the
hard x-ray absorption mechanism at the surface has not been fully
explored. We show, by specific examples, that since Compton
scattering and pair production are two competing processes in
x-ray absorption in the photon energy up to 5 MeV
\cite{Heitler54,Bethe53,Migdal56}, where the upper limit of the
photon energy depends on the Z-number (the cross-section of
pair-production becomes equal to that of Compton scattering at 5
MeV for Pb; it is 15 MeV for Al), Compton scattering leads to a
partial reflection of x-rays at a sharp boundary with a curvature.
Furthermore, we show that since the basic x-ray transmission
equation remains correct to the single-scattering approximation,
the mean free-path of an incident photon in a given material must
be sufficiently large to neglect the effect of multiple
scattering.

A solid cylinder with different holes in size at both ends is
ideally suited for studying the effect of a partial reflection of
x-rays at a sharp boundary on the absorption coefficients. A
single exposure of an x-ray beam to the cylinder provides a data
for a solid cylinder as well as for a cylinder with a void.
Moreover, the circular curvature allows one to study the angular
dependence of scattered photons at the surface, where the
scattering centers are located.

An analytic solution to Abel's integral equation that is least
sensitive to local random errors is the starting point of our
study. We show, by an analytic method, that Abel's integral
equation admits a different form of a solution than that obtained
by Volterra \cite{Horman35}. This modified form of a solution not
only reduces the computing time by an order of magnitude, but also
allows one to analyze the effect of intensity fluctuations on the
absorption coefficients.

First we review the basic x-ray absorption processes in a high-Z
metallic test object and discuss the relative importance of the
processes at a given energy of an x-ray. Next we review Abel's
integral equation and derive its solutions in Sec. III. We then
study numerical methods of data analysis based on the new solution
to Abel's integral equation in  Sec. IV and compare the numerical
results with the known absorption coefficients of material. The
application of the numerical code is discussed in Sec. V.  In Sec.
VI we discuss an optimal experimental condition for x-ray
diagnostics and the basic requirements of an x-ray source for the
condition. Lastly in Sec. VII, the feasibility of quantitative
x-ray diagnostics is discussed.

\section{\label{sec:level1} X-ray Absorption and Transport
in High-Z Material}

Pulsed high-energy x-ray sources ({\it e.g.} PHERMEX) have
provided a way to obtain x-ray snap shots of dynamical processes
such as imploding spherical targets in an ICF program. These snap
shots provide the transmitted x-ray intensity profiles with a
diffuse background. In PHERMEX facility at Los Alamos, the
high-energy $\gamma$-rays are produced by short pulsed high-energy
electron beam. The high-energy electron beam strikes the target
material in which the energy loss of the beam by bremsstrahlung is
fairly uniformly distributed among the secondary photons of all
energies from zero up to the energy of the primary electron itself
(30~MeV), for which the bremsstrahlung process dominates over the
collision loss \cite{Motz59}.

At some electron energy, the radiation loss is equal to the
collision loss. This energy coincides with the critical energy of
the material, a parameter that also plays an important role in the
shower theory of Rossi \cite{Rossi52,Motz59}. Thus at
high-energies a large fraction of the electron energy is converted
to high-energy $\gamma$-rays in the target which in turn may
interact with test object. The theory of high-energy photon
absorption mechanism (in the order of ${\cal E}_{\gamma}\gg
2mc^{2}=1.02$ MeV) has been extensively studied. The basic physics
is fairly easy to understand and is fully described in Refs.
\cite{Rossi52,Heitler54,Bjorken64,Bethe53}.

Next we discuss the radiation absorption process in a test object;
at high-energies, the radiation absorption by pair-production
(electron-positron) dominates over Compton scattering. Hence the
high-energy $\gamma$ rays create an electromagnetic cascade
shower. The photoelectric effect and the multiple Coulomb
scattering of the secondary electrons by atoms perturb the cascade
shower. This coupled with Compton scattering creates a diffuse
background in a radiographic film. In a thick test object, the
energy of a low-energy secondary electron in the beam is
dissipated by either excitation or the ionization of atoms. We
will discuss further the basic processes in the radiation
absorption below.

In general, as a beam of x-rays penetrates an absorbing material,
its intensity shows an exponential decay. It is because a photon
in the beam is removed from its original direction of penetration
by either absorption or scattering in {\it a single event}. Since
the number of photons removed in a distance $dx$ is proportional
to the beam intensity $I(x)$, the intensity of a monochromatic
beam passing through the matter of a test object must decrease
exponentially and is given by

\begin{equation}
I(x)=I_{0}\exp[-\!\!\int_{x_{0}}^{x}\mu(x)dx],
\label{absorption}
\end{equation}
where the absorption coefficient $\mu$ represents the average
number of absorption and scattering processes a single photon
undergoes per $cm$ in its path and $I_{0}$ is the beam intensity
at $x=x_{0}$.

Neglecting the effect of multiple scattering and photon cascade,
the cross-section for the primary processes ({\it i.e.,}
photoelectric effect, Compton scattering, and pair production) by
which x-rays or $\gamma$-rays are absorbed is given by the sum:
$\sigma_{total}=\sigma_{photo}+Z\sigma_{Comp}+\sigma_{pair}$. The
absorption coefficient can be defined as
$\mu=N\sigma_{total}(cm^{-1})$, where $N$ is the number of atoms
of the absorbing matter per cubic centimeter.

\section{\label{sec:level1} Basic Processes in Radiation Absorption}

Briefly we now digress to summarize the salient points of the
photon absorption processes in condensed matter for the
application of radiography:

\subsection{\label{sec:level2}Photoelectric Effect}

At Photon energies less than 0.5 ~MeV, x-ray interaction with the
material is dominated by photo-absorption in which an x-ray photon
is completely absorbed by a bound electron in atoms. Consequently
the bound electron jumps from its ground state to an excited state
or into the continuum; the latter process is an ionization of the
atom, which is usually more probable than the excitation. The
cross-section for photoelectric effect is derived first by Heitler
\cite{Heitler54}. When the photon energy is not too close to the
absorption edge, the cross-section for the photo-emission of a
single electron from the K shell is given by
$\sigma_{photo}=\sigma_{th}4\sqrt{2}\alpha^{4}Z^{5}(mc^{2}/{\cal
E}_{\gamma})^{7/2}$, where $\sigma_{th}$ is the cross-section
Thompson scattering, $\alpha$ the fine-structure constant, Z the
atomic number, and ${\cal E}_{\gamma}$ the photon energy. Thus in
the region of photon energies of the order of a few KeV, the
cross-section for the photoelectric effect decreases rapidly as
${\cal E}^{-7/2}$.

For high-energy photons, ${\cal E}\gg mc^{2}$, the cross-section
for the photoelectric effect is given by
$\sigma_{photo}=\sigma_{th}(3/2)\alpha^{4}Z^{5}(mc^{2}/{\cal
E}_{\gamma})$. This shows that at high-energies the cross-section
decreases as $1 /{\cal E}_{\gamma}$ but photoelectric effect can
still contribute to x-ray absorption for high-Z material up to 5
MeV ({\ e.g.,} Pb) although its effect becomes increasingly small.
At Photon energies higher than 1 MeV, the cross-section for the
photo-absorption becomes negligible, however, compared to that of
Compton scattering.

\subsection{\label{sec:level2}Compton Scattering}

Scattering of photons by free elctrons plays an important role in
x-ray or $\gamma$-ray absorption in condensed matter. The
cross-section for Compton scattering by relativistic corrections
was first derived by Klein and Nishina \cite{Bjorken64}. We note
that since the binding energy of the K shell electrons in atoms is
of the order of KeV, the scattering a high energy photon in MeV by
a bound electron is essentially equivalent to Compton scattering
by a free electron.

For a cylindrical test object, the scattering center is well
defined at the surface and the angular dependence of scattered
photons by the surface can be studied by the Klein-Nishina
formula. For high-energy photons, the formula shows that  the
angular distribution of scattered photons tends to peak in the
forward direction, which agrees well with the observation.

The conservation of energy and momentum give
$k/k^{\prime}=1+(k/m)(1-\cos\theta)$, where $k$ and $k^{\prime}$
are the energy of the incident and scattered photons respectively,
and $\theta$ the angle in the units of $\hbar=c=1$, $c$ the
velocity of light.

If the energy of an incident photon is large compared with the
rest mass of an electron, the effect of Compton scattering can be
studied in two extreme cases: for a very small scattering angle,
the change of photon energy is negligible, and at large
scattering angles, the scattered photon energy is always of the
order of $mc^{2}$. Thus the effect of Compton scattering in this
case reduces to the classical Thompson scattering and the change
in the photon energy becomes negligible.

An important point to emphasize is that, at high energies, the
cross-section for Compton scattering $\sigma_{Comp}$ varies as
$(\pi\alpha^{2}/km)[\ln(2k/m)+1/2]$, where $\alpha/m$ is the
classical electron radius and $k$ is the energy of a photon in
the units of $\hbar=c=1$. Thus the cross-section for Compton
scattering decreases with the increasing energy of an incident
photon. This is precisely the reason why the penetrating power of
x-rays (or $\gamma$-rays) increases with increasing energy so
long as no other absorption processes are important.

\subsection{\label{sec:level2}Pair Production}

In the Coulomb field of a nucleus, a high-energy photon creates an
electron-positron pair ({\i.e., } $\gamma\rightarrow
e^{-}+e^{+})$. No pair production can occur if the photon energy
is less than $2mc^{2}$ or in the absence of the nuclear Coulomb
field required for the conservation of energy and momentum
\cite{Migdal56,Bjorken64,Motz69}. The cross-section for the
pair-production increases rapidly with the photon energy and
reaches an asymptotic value in the relativistic limit. If the
photon energy ${\cal E}_{\gamma}\gg (mc^{2}/\alpha)Z^{-1/3}$, then
the cross-section takes the asymptotic form: $\sigma_{pair}=
(\alpha^{3}/m^{2})Z^{2}[(28/9)\ln(183/Z^{1/3})-(2/27)]$, where the
effect of atomic electrons through their screening of the nuclear
charge is taken into account. This limiting form shows that the
cross-section increases with the square of the Z-number. We also
note that, for high energy photons, most of the electrons and
positrons are emitted in the forward direction.

Since both the photoelectric effect and Compton scattering
decrease as the photon energy increases, the pair production is
the dominant absorption mechanism for very high-energy photons.
Furthermore the critical energy above which pair production
dominates in photon absorption decreases with increasing Z number.
For example, the cross-section for pair production becomes equal
to that of  Compton scattering at about 5~MeV for Pb.

In elementary quantum field theory, there is a useful, general
substitution rule which is valid to arbitrary orders and which
relates processes of the type, $A+B \rightarrow C+D$, to the
processes $ A+\bar{C} \rightarrow \bar{B}+D$, where $\bar{B}$
denotes the antiparticle of B. This substitution rule relates
Compton scattering to the annihilation of an electron-positron
pair into two photons ({\it i.e.,} $e^{-}+e^{+}\rightarrow
2\gamma)$ \cite{Motz69}. In a similar manner, the pair production
process in condensed matter is closely related to bremsstrahlung
process by high energy electrons \cite{Migdal56,Motz59}.

\subsection{\label{sec:level2}Multiple Scattering and the Effect of
Photon Cascades}

In condensed matter, $\gamma$-rays can undergo multiple
scatterings, a sequence of scattering of the incident photon by
bound electrons, if the mean free path of the photon is small
compared to the size of a medium, which is usually the case for
high-Z material \cite{Motz59}. Thus the critical parameter
determining the effect of multiple scattering becomes negligible
if the size of a medium is less than a few times of the mean free
path of a photon, which is given as $\lambda_{(mfp)}=\mu^{-1}$,
where $\mu$ is the absorption coefficient at a given energy in
material ({\it e.g.,} $\lambda_{(mfp)}=2.13$~cm at ${\cal
E}_{\gamma}=10 mc^{2}$ for Pb). We notice from Figure 32 of Bethe
and Ashkin \cite{Bethe53} that the absorption coefficient for any
given material goes through a minimum at some energy ${\cal
E}_{\gamma m}$ of an incident photon energy. The nature of the
transmitted x-rays depends on whether the energy of an incident
photon is greater or less than  ${\cal E}_{\gamma min}$. For
example, if the energy of a primary photon is less than ${\cal
E}_{\gamma min}$, the primary photon is more penetrating than the
secondary photon. Thus in a test object whose dimension is about
the same as the mean free path of a photon, the effect of multiple
scattering is negligible.

It would be difficult to carry out a systematic treatment of the
multiple scattering involving photoelectric effect, Compton
scattering, and pair production. In spite of this complexity, an
elaborate analysis of x-ray transmission in thick targets has
been carried out analytically that shows how to correct the
simple transmission rate Eq. (1), beyond its simplifying
approximations. Fortunately, we now have an excellent Monte Carlo
code (EGS4) \cite{EGS4} developed at SLAC, which can be used to
test its accuracy against the experimental data. Furthermore,
this test helps us understand the detailed absorption and
production mechanism of $\gamma$-rays for a given electron beam
from the accelerator, such as PHERMEX facility at Los Alamos.

At high-energies $5 MeV \geq {\cal E}_{\gamma m}\geq 1 MeV$, where
absorption by Compton scattering dominates, one may treat the
multiple scattering as a series of small angle Compton scattering
in which the photon energy loss is small for $\gamma$-rays that
penetrate a large distance in the absorbing medium. With this
approximation, the asymptotic solution for the transmission rate
was obtained. It contains a factor $\exp^{-\mu_{m}x}$, a build-up
factor, where $\mu_{m}$ is the smallest absorption coefficient for
a given material ({\it e.g.,} $\mu_{m}$ is the absorption
coefficient at 2.5~MeV for Pb). The build-up factor depends on
whether the incident photon energy is greater or smaller than the
energy at which the absorption coefficient is minimum
\cite{Bethe53,Hirschfelder48}.

It is also well known that at very high energies (${\cal {E}}\gg
10 mc^{2}$), absorption by pair production dominates and thus the
secondary Compton scattered photons may become more penetrating.
Under this condition, the spectral equilibrium is controlled by
the hardest secondary photons and the transmission takes a form
$x^{-5/6}e^{-\mu_{m}x+bx^{1/2}}$, where $\mu_{m}=\mu(\lambda_{m})$
is the minimum value at some wavelength (or energy) of an incident
photon for any given material
\cite{Wick49,Karr49,Spencer49,Fano50}. Here the energy of an x-ray
photon is related to its wavelength by ${\cal E}_{\gamma
min}=hc/\lambda$, where $h$ is Planck's constant. In laboratory
units, $\lambda=12.4\times10^{-3}/{\cal E}_{\gamma min}(MeV)$,
where $\lambda$ is the wavelength in angstrom.

At extremely high energies, x-ray absorption by pair production
becomes increasingly important. In this process, a photon creates
an electron-positron pair which in turn can produce secondary
photons by bremsstrahlung processes; the pair can also annihilate
into two photons ({\i.e.,} $e^{-}+e^{+}\rightarrow2\gamma$). The
secondary photons can also undergo photo-absorption, Compton
scattering or pair production, a phenomenon known as a photon
cascade. The effects on the absorption coefficient of multiple
scattering and photon cascades can be appreciable, depending on
the range of an incident photon energy \cite{Rossi52,Rossi41,
Tsai74}.

\subsection{\label{sec:level2}Reflection of X-Rays at a Sharp
Boundary}

In connection with Compton scattering of soft photons, we discuss
the reflection of an x-ray beam at a sharp boundary by treating
the x-ray beam as an incident electromagnetic wave. A
semi-classical analysis of the reflection of x-rays gives the
index of refraction for x-rays as $n=1-\delta-i\beta$, where
$\beta=\lambda\mu/(4\pi)$ and
$\delta=1/(4\pi)(e^{2})/(mc^{2})[(N_{0}\rho)/A][Z+\Delta
f^{'}\lambda^{2}]$. Here $N_{0}$ is the Avogadro's number, $\rho$
the density, and $A$ the atomic number weight, $\Delta f^{'}$ is
small away from the absorption edges \cite{Parratt54}. Since the
index of refraction is less than 1, the reflection can occur if
the incident glancing angle is sufficiently small.

From Snell's law, the critical angle below which the total
reflection can occur is determined by $\phi=\sqrt
2\delta\simeq\rho^{1/2}_{e}\lambda$, where we have neglected the
absorption term. This semi-classical analysis is often useful in
identifying the reflection of soft x-rays from a bremsstrahlung
source at a sharp boundary. For an x-ray wavelength of $1.03
\times10^{-2} \AA$, the critical angle
$\phi_{cr}\simeq5.8\times10^{-5} rad$. As the incident glancing
angle increases from the critical angle, the reflection intensity
decreases.

\subsection{\label{sec:level2}Fluctuations of X-Ray Beam Intensity}

An important refinement of the x-ray absorption process can
be made from the observation that in a disordered  medium the
transmitted beam intensity shows a local fluctuation. Given the
enormous amount of effort that has been devoted to the x-ray
absorption mechanism, remarkably little is known about the nature
of intensity fluctuations. The concept of localization of
electrons in a solid was invoked in 1958 by Anderson to explain
the electron transport phenomena in the presence of impurities
\cite{Anderson58}. The concept is usually associated with a
disordered medium in which interference of multiple scattering of
electrons by impurities give rise to enhanced backscattering.

We can see from Eq.~(\ref{absorption}) that apart from the $K$ and
$L$ absorption edges the total cross section of the primary
processes is a smooth function of an incident photon energy. The
transmission rate Eq.~(\ref{absorption}) was derived based the
single scattering approximation, which is valid as long as the
mean free path of an incident photon for any given material is not
small compared with the dimension of a test object. This implies
that an independent, identical N atom model cannot explain the
fluctuations of beam intensity. One final comment on the
fluctuation of a transmitted x-ray intensity is note-worthy.
Anderson localization was observed in experiments with a
monochromatic light source \cite{Alvada85,Wolf85}. Unfortunately,
the wide spectrum of a bremsstrahlung source can be a contributing
factor to the intensity fluctuations but it is difficult to assess
this effect. We can test our qualitative analysis of beam
transportation in a material in each step by using the code EGS4
\cite{EGS4} which incorporates many different absorption processes
which are almost impossible to carry out a detailed analysis in
practice.

\section{\label{sec:level1} Abel's Integral Equation and Its Solution}

 We now consider a symmetric cylinder whose material
composition is uniform in the $z$-direction perpendicular to the
incident x-ray beam and focus on an infinitely thin disk of the
cylinder that is sufficiently  far from the x-ray source. This
allows one treat the x-ray source as a spatially uniform beam,
which is an essential requirement for the derivation of Abel's
integral equation \cite{comment3}. Furthermore we assume that a
detector accepts only those x-rays which traverse a narrow slice
from the source ({\it i.e.,} a narrow beam approximation).

\subsection{\label{sec:level2} Abel's Integra Equation}

\begin{figure}[tbp]
\includegraphics[bb=0in 0in 6in 4in,totalheight=4in,height=4in,width=6in,
keepaspectratio=true]{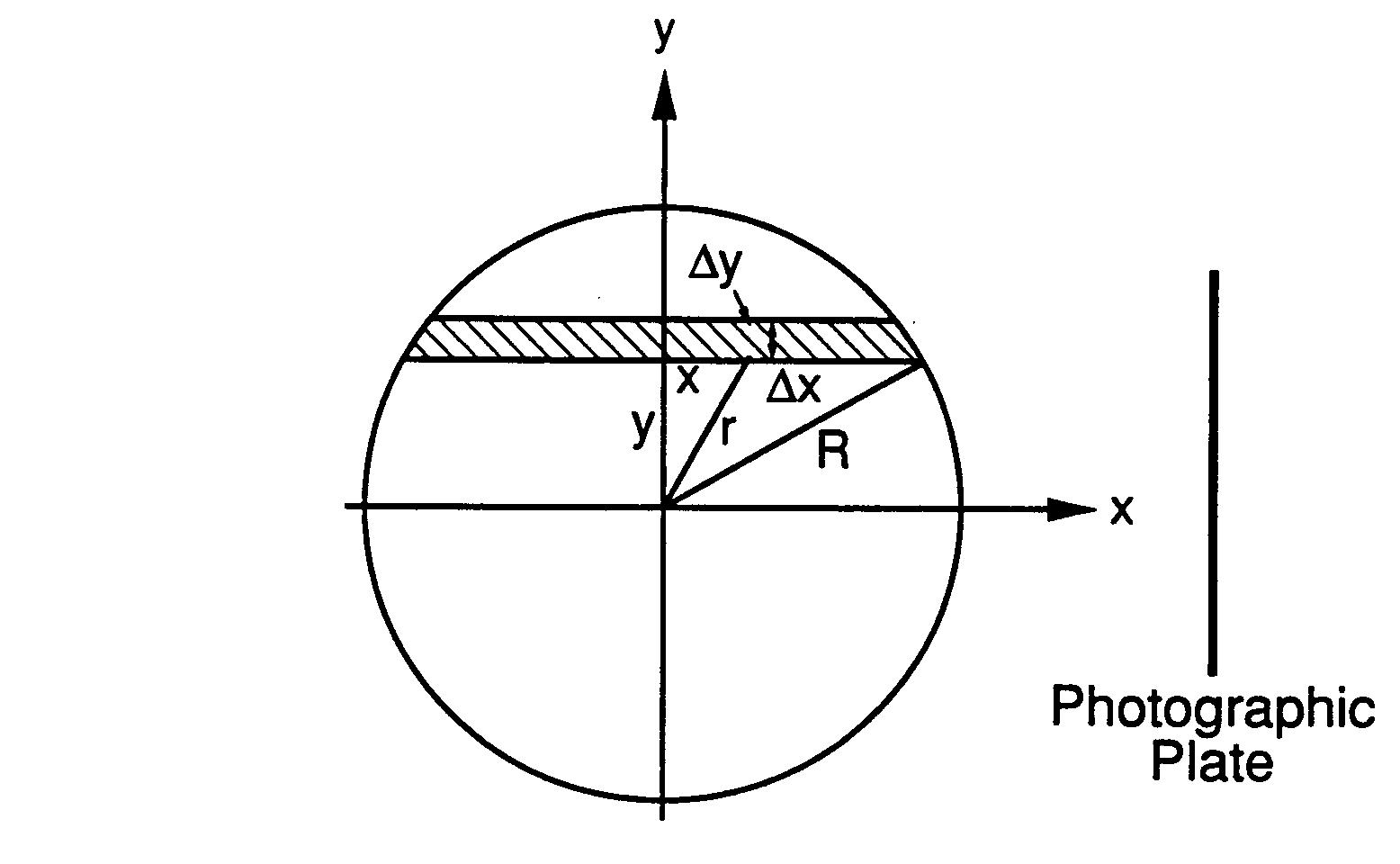} \vspace{0.0truecm}
\caption{The coordinate system for x-ray transmission
measurements. x is the x-ray beam axis, y is the distance of a
thin strip, which is parallel to the x-axis, from the center, and
the cylinder is assumed to be translational invariant with respect
to the z-axis.} \label{fig:fig2}
\end{figure}

For a pure absorption or a single scattering, the flux of an
x-ray beam emerging from the disk, $I(y)$ along the slice is
given by
\begin{equation}
I(y)\delta y\delta z=I_{0}\delta y\delta z\:
exp[-\int_{x_{0}}^{x_{0}}\mu(r)dx],\
\label{flux}
\end{equation}
where $I_{0}$ is the flux of an incident monochromatic beam [see
Figure \ref{fig:fig1}] and is assumed to be independent of $y$,
since we assume a uniform beam as an x-ray source. In Figure
\ref{fig:fig2}, $r^{2}=x^{2}+y^{2}$ and $x_{0}^{2}=R^{2}-y^{2}$,
where $R$ is the radius of the disk. We have also assumed the
absorption coefficient $\mu(r)$ is cylindrically symmetric and
hence our analysis is applicable to an one-dimensional problem,
which is often misunderstood in practice.

Before displaying Abel's integral equation, which is the
principal object of this paper, it is appropriate to comment on
the narrow beam approximation. In an ideal situation, the
transmitted x-ray intensity $I(y)$ must remain nearly
monochromatic as the incident x-ray beam, since the scattered
x-rays are assumed not to reach a detector ({\it i.e.,} a narrow
beam approximation). In the presence of Compton scattering the
narrow beam approximation breaks down if the scattering angle is
large. This is the reason why we observe a diffuse background in
an x-ray film.

If we define $f(y)=\ln(I_{0}/I)$, then Eq.~(\ref{flux}) becomes
\begin{equation}
f(y)=2\int_{0}^{x_{0}}\mu(r)dx \label{abs2}
\end{equation}
Here $f(y)$ gives the sum of absorption or scattering along the
slice. This quantity appears naturally in the integral equation
for all observable which, like an x-ray absorption, represents
strictly local interactions of x-rays with atoms. We shall
therefore give a brief discussion of $f(y)$ itself, before
returning to the details of Abel's integral equation.

In slab geometry, $f(y)$ remains positive or become zero if there
is no absorption. In the presence of Compton scattering, $f(y)$
tends to give a value less than the expected value because of
smear from the adjacent slices. Although the effect of Compton
scattering is negligible in slab geometry ({\it i.e.,} a white
noise), its effect becomes serious in a test object with a
geometrical structure; $f(y)$ can give a negative value which
signals a breakdown of a narrow beam approximation. Isolating
$f(y)$ from the background in an x-ray transmission then requires
the subtraction of the structure-dependent quantity, which is
unknown \textit {a priori}, a process requiring very accurate
data analysis.

Let us now return to the derivation of the integral equation. If
we eliminate the variable $x$ in favor of $r$ [see Figure
\ref{fig:fig2}], Eq.~(\ref{abs2}) can be written as
\begin{equation}
f(y)=2\int_{y}^{R}\frac{\mu(r)r dr}{(r^{2}-y^{2})^{1/2}}.
\label{abel}
\end{equation}
This is a form of Abel's integral equation \cite{Whittaker53}.

 \subsection{\label{sec:level2} Solutions to Abel's Integral Equation}

With Abel's transformation, Volterra's solution to
Eq.~(\ref{abel}) is given as \cite{Horman35}
\begin{equation}
\mu(r)=-\frac{1}{\pi}\int_{r}^{R}\frac{f^{'}(y)dy}{(y^{2}-r^{2})^{1/2}},
\label{abel2}
\end{equation}
where the prime on $f(y)$ denotes a differentiation with respect
to its argument.

Equation (\ref{abel}) is our fundamental expression for the
material absorption coefficient. Its equivalence to the x-ray
emission from a hot plasma is clear from the derivation of
Eq.~(\ref{abel}) but can be made explicit by replacing $I(y)
\leftrightarrow I_{0}(y)$ and $\mu(r) \leftrightarrow e(r)$ in
Abel's integral equation. Thus Eq.~(\ref{abel2}) is applicable to
both the radiation absorption and emission processes
\cite{Horman35,Maecker54,Friedrich59}.

In practice, $f(y)$ is given as a set of discrete data taken from
a detector or an x-ray film rather than as an analytic function.
Normally, experimental data also contain a diffuse background
({\it i.e.,} random errors) due to Compton scattering or photon
cascades. In the presence of fluctuations of a beam intensity, it
is difficult to eliminate these errors either by
Fourier-filtering or other numerical methods. The differentiation
on the data points introduces spikes in noise in the integral.
Thus a numerical representation of the data poses a difficult
problem that must be treated with care. The numerical integration
of Eq.~(\ref{abel2}) is best avoided entirely, if the numerical
accuracy is desired.

Bockasten \cite{Bockasten61}, and recently Wing \cite{Wing84}
also recognized this point, but their calculations of polynomial
fits of data points or expansion of the kernel of integral
equation limit the accuracy of the results. As an alternative to
the use of Eq.~(\ref{abel2}), Barr \cite{Barr62} obtained a
solution of Eq.~(\ref{abel}) by means of Abel's transformation.
His solution allows one to devise an efficient smoothing method
for data having a small random errors. His numerical algorithm
eliminates the random errors after the integration, but before
differentiation. This procedure ensures smooth results for a
Gaussian type of  density profiles. However, his solution is
numerically unstable near the center. To overcome this difficulty
we look for a different solution to Eq.~(\ref{abel}) that is
numerically more stable.

We start with the identity \cite{Whittaker53},
\begin{equation}
\frac{\pi}{\sin\mu\pi}=\int_{0}^{\infty}\frac{y^{\mu-1}}{1+y}dy,
\label{identity}
\end{equation}
where $0<\mu<1$.

By making a transformation $y=(z^{2}-x^{2})/(x^{2}-\xi^{2})$, we
obtain
\begin{equation}
\frac{\pi}{\sin\mu\pi}=2\int_{\xi}^{z}\frac{x}{(z^{2}-x^{2})^{1-\mu}(x^{2}-\xi^{2})^{\mu}}dx.
\label{abel3}
\end{equation}

For $\mu=1/2$, Eq.~(\ref{abel3}) reduces to
\begin{equation}
\frac{\pi}{2}=\int_{\xi}^{z}\frac{x}{(z^{2}-x^{2})^{1/2}(x^{2}-\xi^{2})^{1/2}}dx.
\label{abel4}
\end{equation}

Next we form a product integral
\begin{subequations}
\label{abel5}
\begin{eqnarray}
F(z) &=& \int_{z}^{R}\frac{y f(y)dy}{(y^{2}-z^{2})^{1/2}} \nn\\
     &=&2\int_{z}^{R}dy \int_{y}^{R}\frac{r\mu(r)y
dr}{[(y^{2}-z^{2})(r^{2}-y^{2})]^{1/2}}\\
     &=&\int_{z}^{R}r\mu(r)dr\left[\int_{z}^{r}\frac{y
dy}{[(y^{2}-z^{2})(r^{2}-y^{2})]^{1/2}}\right]\\
     &=&\pi\int_{z}^{R}r\mu(r)dr, \label{abel6}
\end{eqnarray}
\end{subequations}
where we have used Dirichlet's formula \cite{Whittaker53} and
Eq.~(\ref{abel4}).

Differentiating both sides of Eq.~(\ref{abel6}), we obtain
\begin{equation}
z\mu(z)=-\frac{1}{\pi}\frac{\partial}{\partial
z}\int_{z}^{R}\frac{yf(y)dy}{(y^{2}-z^{2})^{1/2}},
\end{equation}
which can be rewritten as
\begin{equation}
\label{abel7} \mu(r)=-\frac{1}{\pi r}\frac{\partial}{\partial
r}\int_{r}^{R}\frac{yf(y)dy}{(y^{2}-r^{2})^{1/2}}.
\end{equation}
Here we notice that, in the limit $r \rightarrow 0$, $\mu(r)$
becomes numerically unstable, since $\mu(r)$ is indeterminate at
the limit \cite{Barr62}.

We now derive a different form of a solution to Abel's integral
equation Eq.~(\ref{abel}), which is equivalent to
Eq.~(\ref{abel2}) but is less sensitive to random errors, and
removes the difficulty in Barr's solution Eq.~(\ref{abel7}).

As before, we form a product integral

\begin{subequations}
\begin{eqnarray}
G(z) &=&
\int_{z}^{R}\frac{z}{y}\frac{f(y)dy}{(y^{2}-z^{2})^{1/2}}\\
&=&2\int_{z}^{R}dy\left[\frac{z}{y}\frac{1}{(y^{2}-z^{2})^{1/2}}\right]
\int_{y}^{R}\frac{\mu(r)r dr}{(r^{2}-y^{2})^{1/2}}. \label{new}
\end{eqnarray}
\end{subequations}

By means of Dirchlet's formula \cite{Whittaker53}, the integral
can be rearranged as

\begin{equation}
G(z)=2\int_{z}^{R}dr\mu(r)\left[\int_{z}^{r}\frac{rzdy}{y[(r^{2}-y^{2})
(y^{2}-z^{2})]^{1/2}}\right].
\end{equation}

With the transformation
$[(r^{2}-y^{2})(y^{2}-z^{2})]^{1/2}=\eta(r^{2}-y^{2})$. the
integral in the bracket $[~~~]$ can be evaluated by an elementary
method, yielding
\begin{equation}
G(z)=\pi\int_{z}^{R}\mu(r)dr.
\end{equation}

Thus we have
\begin{equation}
\int_{z}^{R}\frac{z}{y}\frac{f(y)dy}{(y^{2}-z^{2})^{1/2}}=\pi\int^{R}_{z}\mu(r)dr.
\label{new1}
\end{equation}

Using Eq~(\ref{new1}) we derive the following equation by taking
the advantage of the near continuous nature of the function
$f(y)$ and with the similar transformation:

\begin{subequations}
\label{last}
\begin{eqnarray}
\mu(r) &=&-\frac{d}{dr}{\cal F}(r),\\
\label{int2}
{\cal F}(r) &\equiv&
\frac{1}{\pi}\int_{r/R}^{1}\frac{f(r/z)dz}{(1 -z^{2})^{1/2}}.
\end{eqnarray}
\end{subequations}

Some features of Eqs.~(\ref{last}) are similar to those of Barr,
Eq.~(\ref{abel7}) \cite{Barr62}. While he starts with Abel's
transformation, we start, however, from Dirichlet's formula
\cite{Whittaker53} and carry out the final integral by an
elementary method. An important point about this approach is that
it avoids the direct differentiation of data points. Instead, we
integrate over the data points, which, with proper choice of mesh
points, smooths random errors in the data.

Next we show that Abel's integral equation admits three different
forms of a solution, which are all equivalent as shown below:
\begin{subequations}
\label{test}
\begin{eqnarray}
F(r)-r G(r) &=& \int_{z}^{R}\frac{y
f(y)dy}{(y^{2}-r^{2})^{1/2}}-\int_{z}^{R}\frac{r^{2}}{y}\frac{f(y)dy}{(y^{2}-r^{2})^{1/2}}\\
            &=&\int_{r}^{R}\frac{(y^{2}-r^{2})^{1/2}}{y}f(y)dy.
\end{eqnarray}
\end{subequations}

Differentiating both sides of Eq.~(\ref{test}) with respect to
$r$, we obtain

\begin{subequations}
\begin{eqnarray}
\frac{d}{dr}F(r)-G(r)-r\frac{d}{dr}G(r) &
=&-\int_{z}^{R}\frac{r}{y}\frac{f(y)dy}{(y^{2}-r^{2})^{1/2}}\\
&=&-G(r). \label{test1}
\end{eqnarray}
\end{subequations}

Thus we have
\begin{equation}
\frac{d}{dr}G(r)=\frac{1}{r}\frac{d}{dr}F(r),
\end{equation}
which shows that the solution by Barr \cite{Barr62} is equivalent
to Eq.~(\ref{new1}).

To show that Eq.~(\ref{abel4}) is equivalent to Volterra's
solution, we integrate $F(r)$ by parts and obtain
\begin{equation}
\label{test2}
F(r)=-\int_{r}^{R}(y^{2}-r^{2})^{1/2}f^{\prime}(y)dy,
\end{equation}
where the boundary condition $f(R)=0$ was used.

Next we differentiate both sides of Eq.~(\ref{test2}) and obtain
\begin{equation}
\frac{\partial}{\partial
r}F(r)=r\int_{r}^{R}\frac{f^{\prime}(y)}{(y^{2}-r^{2})^{1/2}}dy.
\end{equation}

Hence we have
\begin{equation}
\frac{1}{\pi}\int_{r}^{R}\frac{f^{\prime}(y)}{(y^{2}-r^{2})^{1/2}}dy=
\frac{1}{\pi r}\frac{\partial}{\partial
r}\int_{r}^{R}\frac{yf(y)}{(y^{2}-r^{2})^{1/2}}dy,
\end{equation}
which completes the proof that there are indeed three different
solutions to Eq.~(\ref{abel}).

We may choose any one of the three solutions. The simplest
numerical method, in principle, and the one that is least
sensitive to small random errors, and the most computationally
efficient method provide the criteria for selecting a particular
form of the solution: Eq.~(\ref{last}) meets all of these
criteria.

\section{\label{sec:level1}Numerical Methods}

Before we proceed with the numerical computation, it is necessary
to study the analytic structure of Eq.~(\ref{abel}) and
Eq.~(\ref{int2}). We see at once that both  Eq.~(\ref{abel}) and
Eq.~(\ref{int2}) possess an end-point singularity. Therefore, the
numerical accuracy will depend on how we treat the end-point
singularities in these equations.

\subsection{\label{sec:level2}End-Point Singularity}

We may proceed with our numerical calculations by either
polynomial expansion of data points similar to Bockasten's method
\cite{Bockasten61} or by carrying out the numerical integration
for a given set of data by extrapolation (or interpolation).
However, in the presence of fluctuations in the x-ray intensity,
the former approach tends to introduce errors which are too large
to neglect and leads to erroneous results. The latter approach
suffers, of course, from the difficulty of treating the end-point
singularity. Fortunately the numerical algorithm by Bickley
\cite{Jeffereys53} has been very useful for the evaluation of
singular functions at the integral limits.

We carry out the numerical integrations of  Eq.~(\ref{abel}) and
Eq.~(\ref{int2}) using the numerical scheme of Jeffereys and
Jeffereys \cite{Jeffereys53}, which employ Bickley's formula near
the integral limits and carry out the remaining integral by the
Gregory formula. A more elaborate but similar end-point formula
has been also developed by Young \cite{Young54}. However, we
prefer that of Jeffereys and Jeffereys \cite{Jeffereys53},
because it is easier to adapt to different sets of experimental
data.

Next we discuss the steps to reduce the raw experimental data to
properly extracted data for numerical analysis. There are two
aspects in the processing experimental data: a selection of
necessary data points and optimum background removal. The
selection of data points to carry out the numerical integration
and differentiation depends on the numerical algorithm. In most
experiments, data taken from an x-ray film provide a sufficient
number of data points. By three point interpolation or
extrapolation, we set the data at the cell boundaries for the
integrations and carry out the differentiation at the cell
centers. This process effectively smoothes the data in the short
wavelength regime, but it preserves the fluctuation in the long
wavelength regime. In particular, this numerical scheme is useful
for any type of background removal ({\it e.g.,} white noise), and
to make certain that there are no oscillations in the background
whose frequencies are much higher than those of fluctuations in
the data.

\subsection{\label{sec:level2}Numerical Tests of Self-Consistency}

 Eq.~(\ref{abel}) and Eq.~(\ref{int2}) play critical roles in the
present work. To check the self-consistency of the numerical
code, we have calculated $f(y)$ based on a theoretically model
given by Eq.~(\ref{abel}), which are in turn, used to calculate
the absorption coefficients from Eq.~(\ref{int2}).

In the absence of appreciable noise in $f(y)$, the code calculates
the precise profiles of the absorption coefficients (see Figures
\ref{fig:fig5} and \ref{fig:fig6}; and Figures \ref{fig:fig9} and
\ref{fig:fig10}). It should be noticed that Eq.~(\ref{abel}) has
the same analytic structure as Eq.~(\ref{int2}), and that the same
subroutine in the code can be used to compute both $f(y)$ and
${\cal F}(r)$.

The advantage of using the theoretical values $f(y)$ ({\it i.e.,}
instead of a true experimental value in slab geometry) is that
they can be calculated without including effects of wide spectrum
of an x-ray source and Compton scattering of photons with
electrons in a metal. Moreover they can be used to see if a
significant amount of noise is present in the experimental data.
Thus this intermediate step of analyzing the experimental data is
extremely useful for identifying the large angle Compton
scattering effects.

\section{\label{sec:level1}Numerical Calculations of Absorption Coefficients}

\subsection{\label{sec:level2}Experimental Data}

Within the limits of our approximation on the x-ray absorption
mechanism ({\it i.e.,} neglecting the effects of multiple
scattering and photon cascades since these effects are extremely
small in the present experiment), we may calculate the absorption
coefficients from the experimental data. However there are
several other important effects which we must address: first we
have assumed the x-ray source is monochromatic in the derivation
Eq.~(\ref{absorption}) to calculate $f(y)$. This is perhaps the
most serious assumption of the approximations made in the
derivation of Abel's integral equation. This assumption together
with the spatial uniformity of the x-ray beam is essential for
the determination of absorption coefficients for any given
material by means of Abel's inversion, since the absorption
coefficients for any given material depend on the photon energy
of x-ray source. The spatial uniformity requires a point source
of an x-ray which is the major challenge for an accelerator
physicist because the point source requires a pencil beam of a
high energy electron in a high current accelerator ({\it e.g.,}
PHERMEX at Los Alamos) \cite{SJHan1}. Hence the basic assumption
of a monochromatic x-ray source is hard to achieve, since the
bremsstrahlung source has a wide spectrum; the x-ray spectrum
spans the entire energy range from an incident electron beam
energy of an accelerator to zero ({\it e.g.,} $30$ MeV to zero in
PHERMEX). In other words, the bremsstrahlung spectrum is fairly
flat function of photon energy.

\subsubsection{\label{sec:level3} Progressive Filtering}

If the x-ray intensity at a great depth from a bremsstrahlung
source depends on multiple scattering, the beam intensity tends to
reach an equilibrium state ({\it i,.e.,} the progressive formation
and decay of the hardest secondary soft photons). The reason for
this process is that the progressive filtering of an x-ray beam
from a bremsstrahlung source tends to make the attenuation curve
steeper than $e^{-\mu x}$, but the secondary photons from Compton
scattering can compensate for the steepening if the beam
penetrates to a great depth to allow the multiple scattering. For
example, the mean free path of the average photon energy of 1.2~
MeV in $U^{238}$ ia approximately $1.14 cm$, which is roughly
equal to the size of test object, the effect of the progressive
filtering of an x-ray beam by the test object cannot be
compensated for by any simple physical mechanism.

\subsubsection{\label{sec:level3} Effects of a Void}

Next we turn to a discussion of an x-ray transmission in a
material with a void whose radius is much larger than the
coherence length of a disordered medium. The effect of the void on
the absorption coefficients is similar to the surface absorption
in that incident x-rays see a concave surface where a partial
reflection of the x-rays can occur \cite{Parratt54} and the
primary photons may lose its energy by Compton scattering at a
large scattering angle near the surface ({\it i.e.,} a reflection
of x-rays at the sharp boundary). It differs, however, in that the
net effect of progressive filtering and Compton scattering of
photon depends on a geometrical structure factor. The presence of
a sharp boundary leads to multiple reflections of soft photons in
a void such that an assumed free propagation of photons without a
reflection is no longer appropriate, {\it i.e.}, a radiation trap.

Hence the spatial analysis given in Cormack's paper
\cite{Cormack63} requires a proper correction at the interface
between aluminum and Lucite because of a sharp boundary between
them. In addition, the calculation is invalid because of the
ill-defined gamma ray source which should be assumed to be
spatially uniform to make use of Abel's inversion. Although the
spectrum of $Co^{60}$ source is not as wide as bremsstrahlung
source, the gamma ray beam still experience a partial reflection
at a sharp boundary.

\subsubsection{\label{sec:level3} Experimental Determination of
the undesirable Effects}

\begin{figure}[tbp]
\includegraphics[bb=0in 0in 6in 4in,totalheight=4in,height=4in,width=6in,
keepaspectratio=true]{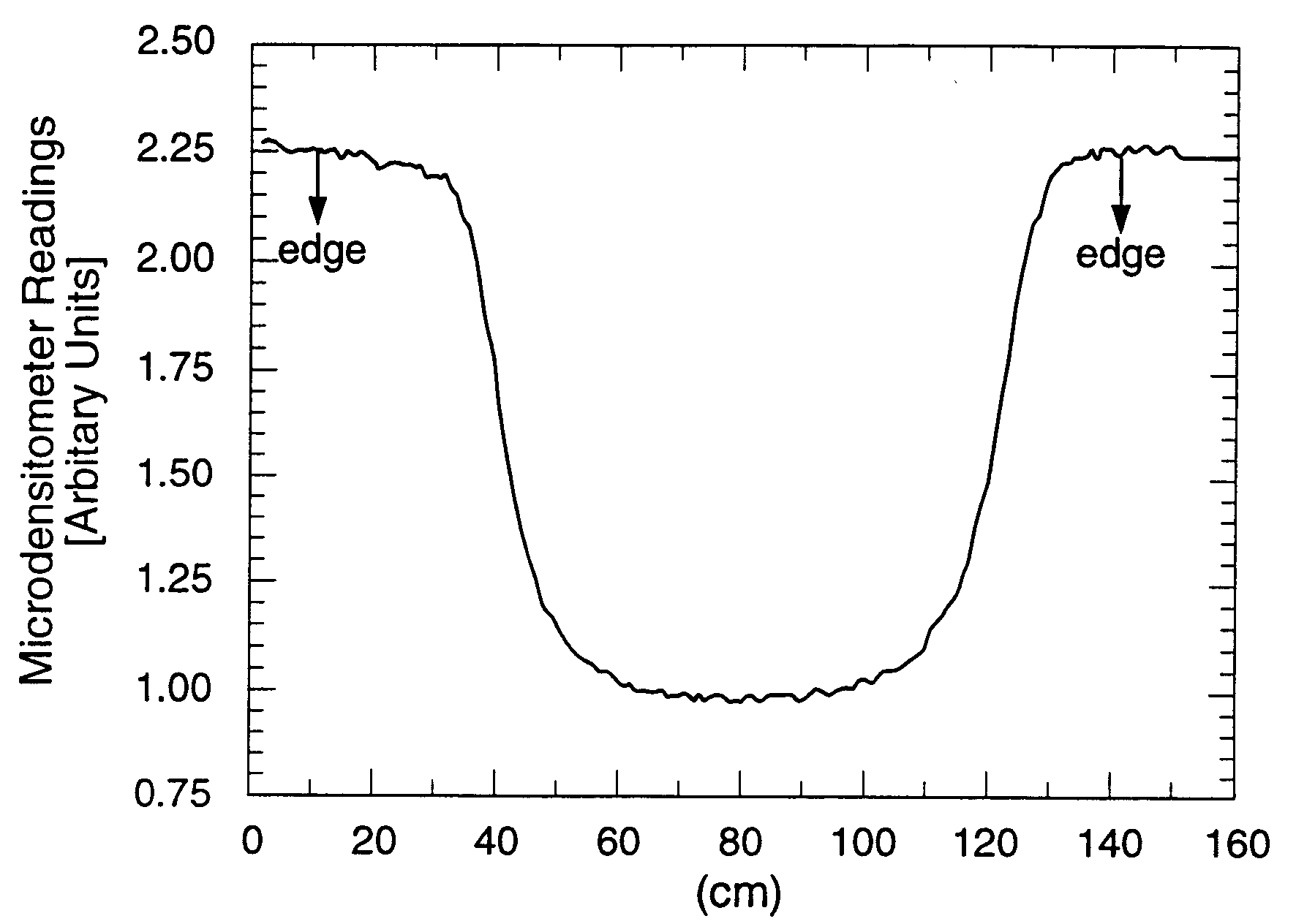} \vspace{0.0truecm} \caption{A
microdensitometer trace of x-ray absorption measurements for a
solid $U^{238}$ cylinder; the x-ray pulse length was approximately
$15$ ns.} \label{fig:fig3}
\end{figure}

\begin{figure}[tbp]
\includegraphics[bb=0in 0in 6in 4in,totalheight=4in,height=4in,width=6in,
keepaspectratio=true]{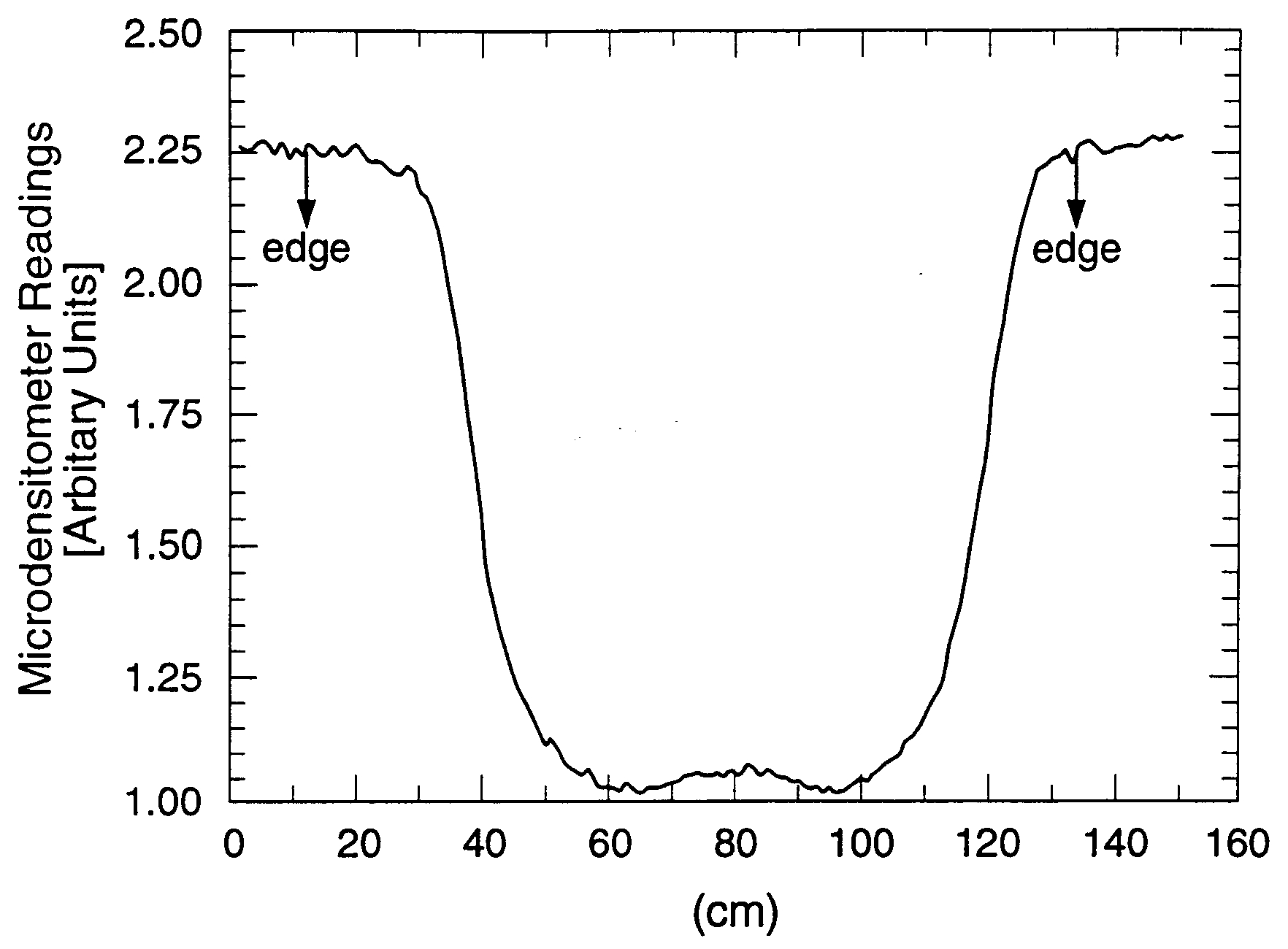} \vspace{0.0truecm} \caption{A
microdensitometer trace of x-ray absorption measurements for a
cylindrical shell of $U^{238}$ with the same x-ray source.}
\label{fig:fig4}
\end{figure}

The present work was undertaken to study the above undesirable
effects experimentally. In particular, to investigate a possible
beam intensity fluctuations, we have selected metallic $U^{238}$
cylinder, which contains a small fraction of impurities that
simulate a random, disordered medium. The schematic picture of the
experimental arrangement is given in Figure \ref{fig:fig1}. The
test object is a cylinder of $U^{238}$, 2~{\it in} long and
0.5~{\it in} diameter with holes on both ends is exposed to a high
energy x-ray source with the average energy of photons ${\cal
E}_{\gamma}\simeq 1.2 MeV$. The x-ray beam emerging from the test
object enters a photographic plate through 0.25~{\it in} thick
lead filter, 3~m~ down stream from the x-ray source. The data from
the x-ray film were then taken, with a microdensitometer
digitized, and then processed by a minicomputer to eliminate the
background. Experimental data sets for both solid and hollow disks
are given in Figures \ref{fig:fig3} and \ref{fig:fig4}, to which a
digital noise reduction technique has not been applied.

\subsection{\label{sec:level2} Theoretical Values Vs Experimental
Data}

\begin{figure}[tbp]
\includegraphics[bb=0in 0in 6in 4in,totalheight=4in,height=4in,width=6in,
keepaspectratio=true]{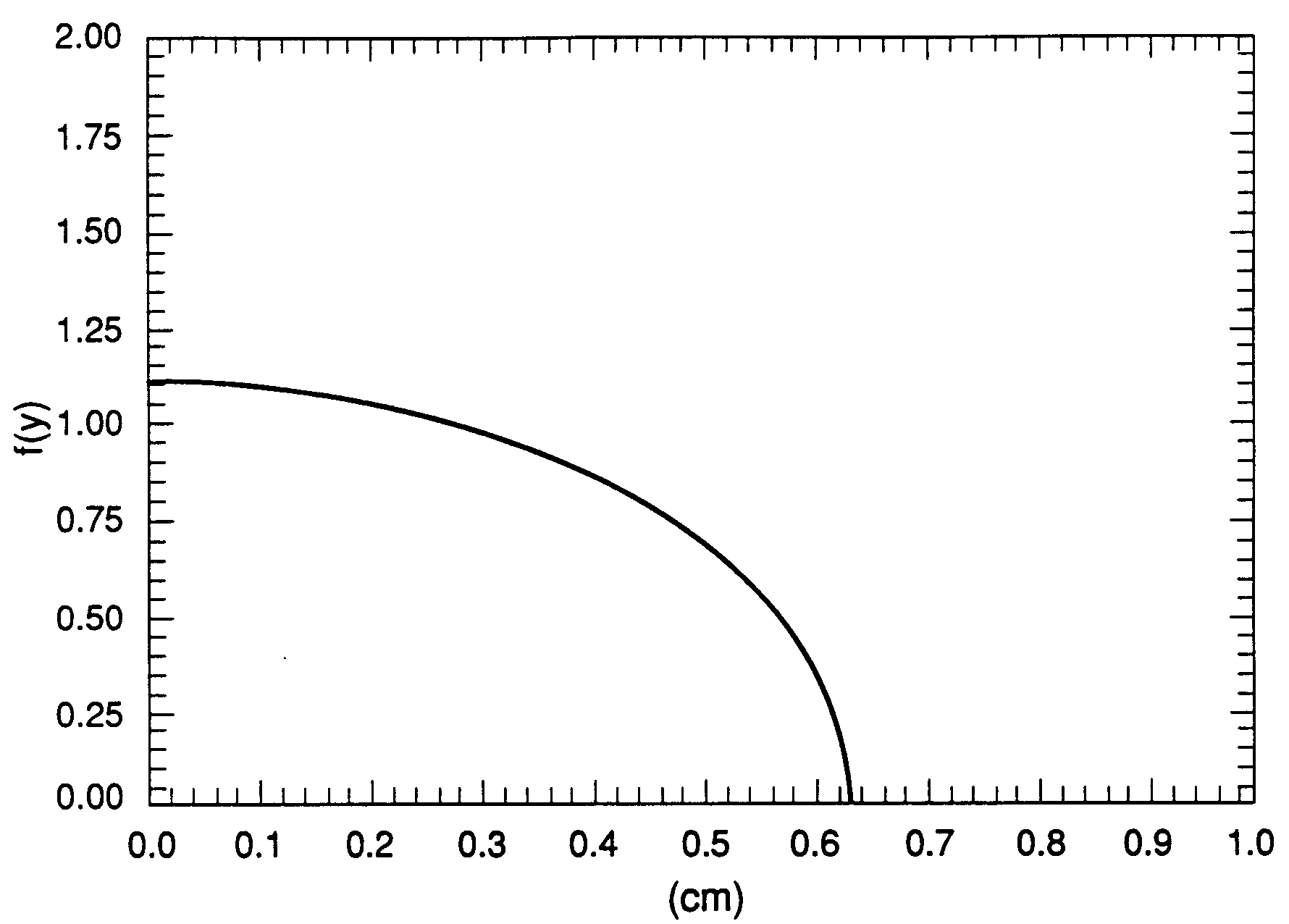} \vspace{0.0truecm}
\caption{The theoretically value $f(y)$ with $\mu=0.875/cm$ for a
solid cylinder. $f(y)$ was computed from Eq.~(\ref{abs2}) with
$R=0.635 cm$. Note that for a constant $\mu$,
$f(y)=2\mu(R^{2}-y^{2})^{1/2}$.} \label{fig:fig5}
\end{figure}

\begin{figure}[tbp]
\includegraphics[bb=0in 0in 6in 4in,totalheight=4in,height=4in,width=6in,
keepaspectratio=true]{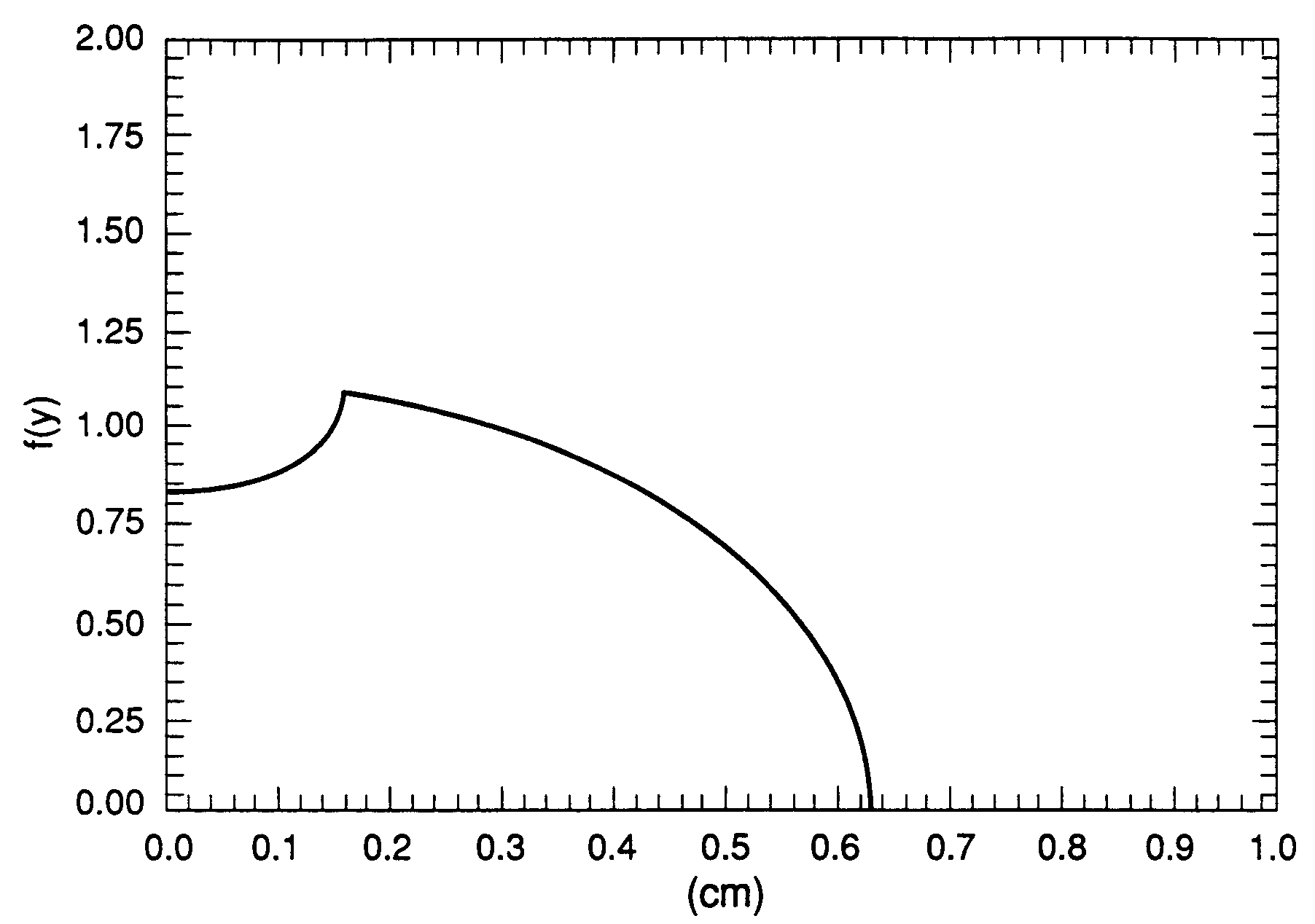} \vspace{0.0truecm}
\caption{The theoretically value $f(y)$ with the same
$\mu=0.875/cm$ for a cylindrical shell with $R=0.635 cm$ and
$r_{1}=0.1588 cm$, where $R$ and $r_{1}$ are the outer and inner
radii of the shell. For a constant $\mu$,
$f(y)=2\mu[(R^{2}-y_{1}^{2})^{1/2}-(r^{2}-y^{2})^{1/2}\theta(y-r_{1})]$,
where $\theta(y-r_{1})=1$ for $y \geq r_{1}$ and
$\theta(y-r_{1})=0$ for $y <r_{1}$ .}
 \label{fig:fig6}
\end{figure}

\begin{figure}[tbp]
\includegraphics[bb=0in 0in 6in 4in,totalheight=4in,height=4in,width=6in,
keepaspectratio=true]{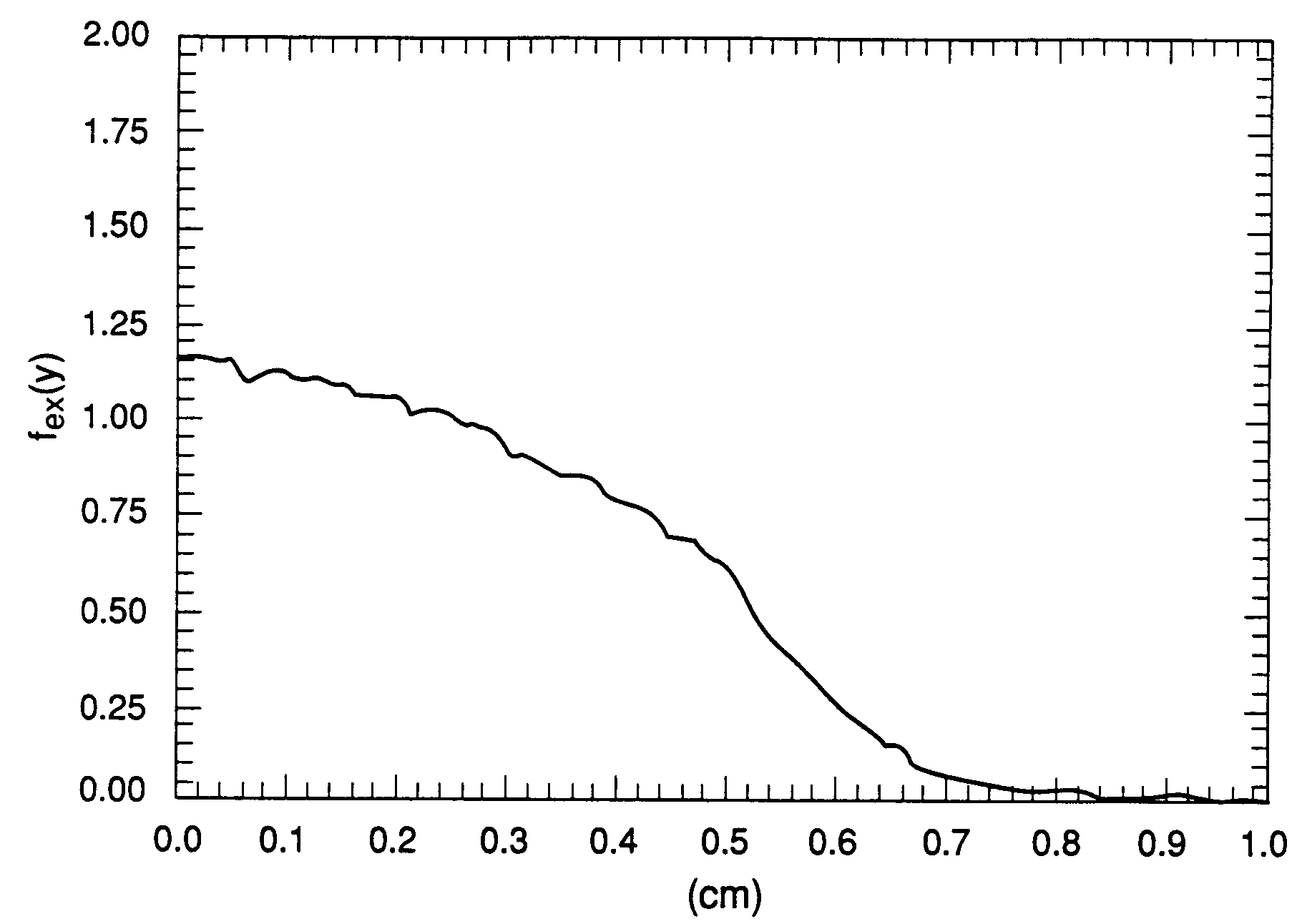} \vspace{0.0truecm}
\caption{The $f_{ex}(y)$ calculated from the experimental data for
a solid $U^{238}$ cylinder; it shows the intensity fluctuation and
a long tail exhibiting a diffuse boundary.} \label{fig:fig7}
\end{figure}

\begin{figure}[tbp]
\includegraphics[bb=0in 0in 6in 4in,totalheight=4in,height=4in,width=6in,
keepaspectratio=true]{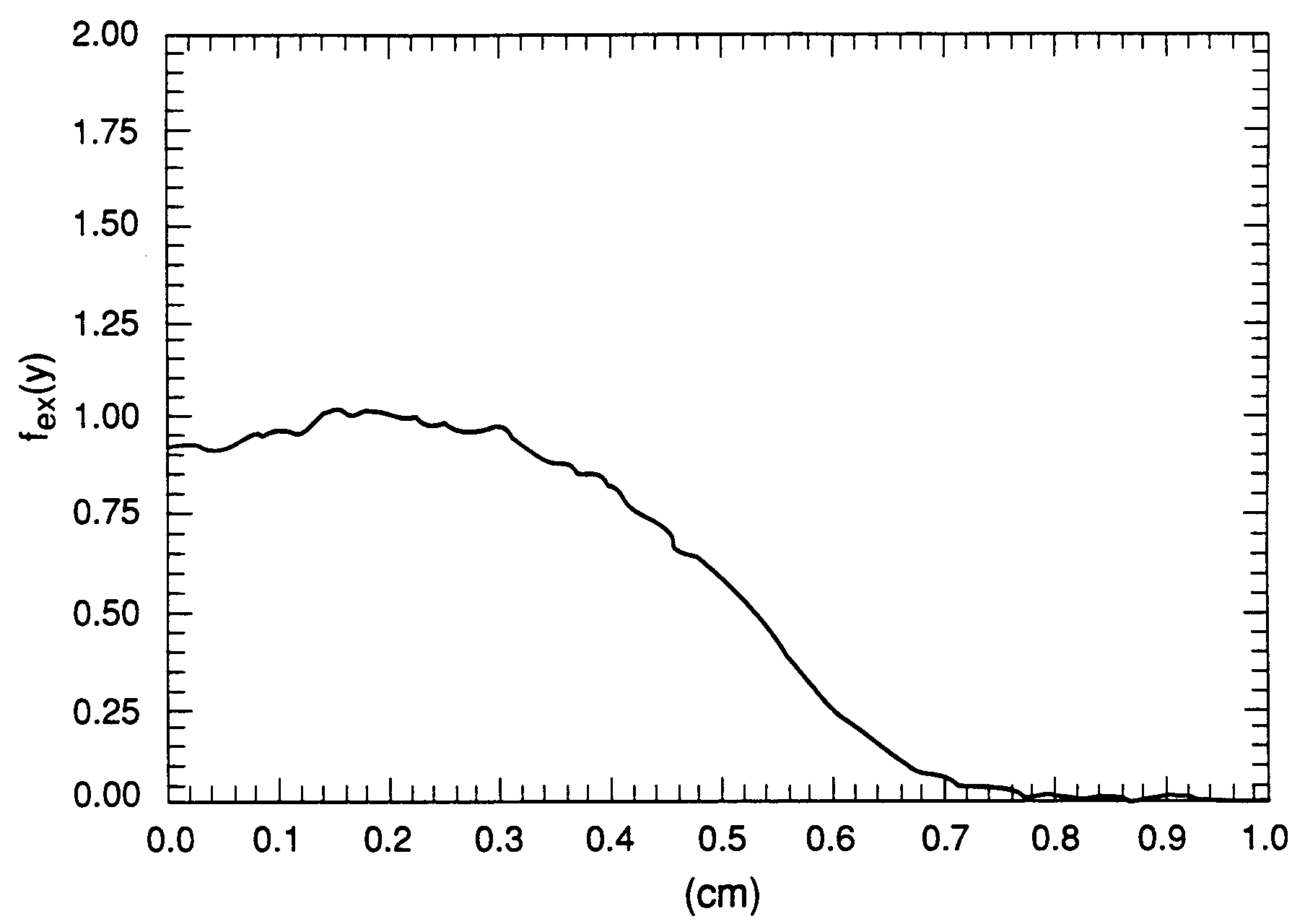} \vspace{0.0truecm}
\caption{The $f_{ex}(y)$ calculated from the experimental data for
a hollow cylindrical (a shell) of $U^{238}$; it shows an excessive
x-ray absorption near the center and a long tail exhibiting a
diffuse boundary.} \label{fig:fig8}
\end{figure}

\begin{figure}[tbp]
\includegraphics[bb=0in 0in 6in 4in,totalheight=4in,height=4in,width=6in,
keepaspectratio=true]{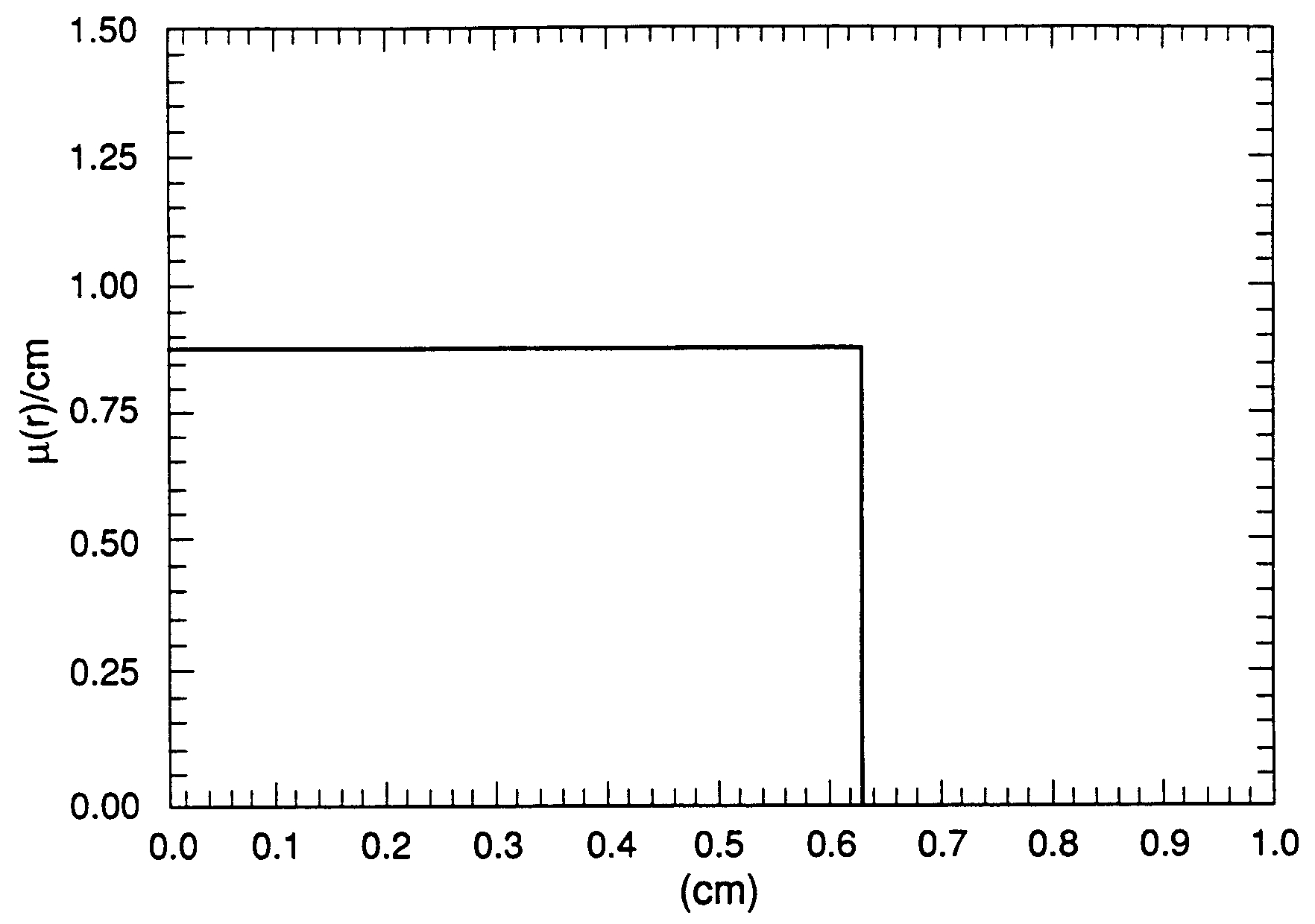} \vspace{0.0truecm}
\caption{The absorption coefficient calculated from
Eq.~(\ref{abel2}) is shown. This demonstrates that the numerical
code gives the anticipated absorption coefficient from the
analytical values given in Figure \ref{fig:fig5}. }
\label{fig:fig9}
\end{figure}

\begin{figure}[tbp]
\includegraphics[bb=0in 0in 6in 4in,totalheight=4in,height=4in,width=6in,
keepaspectratio=true]{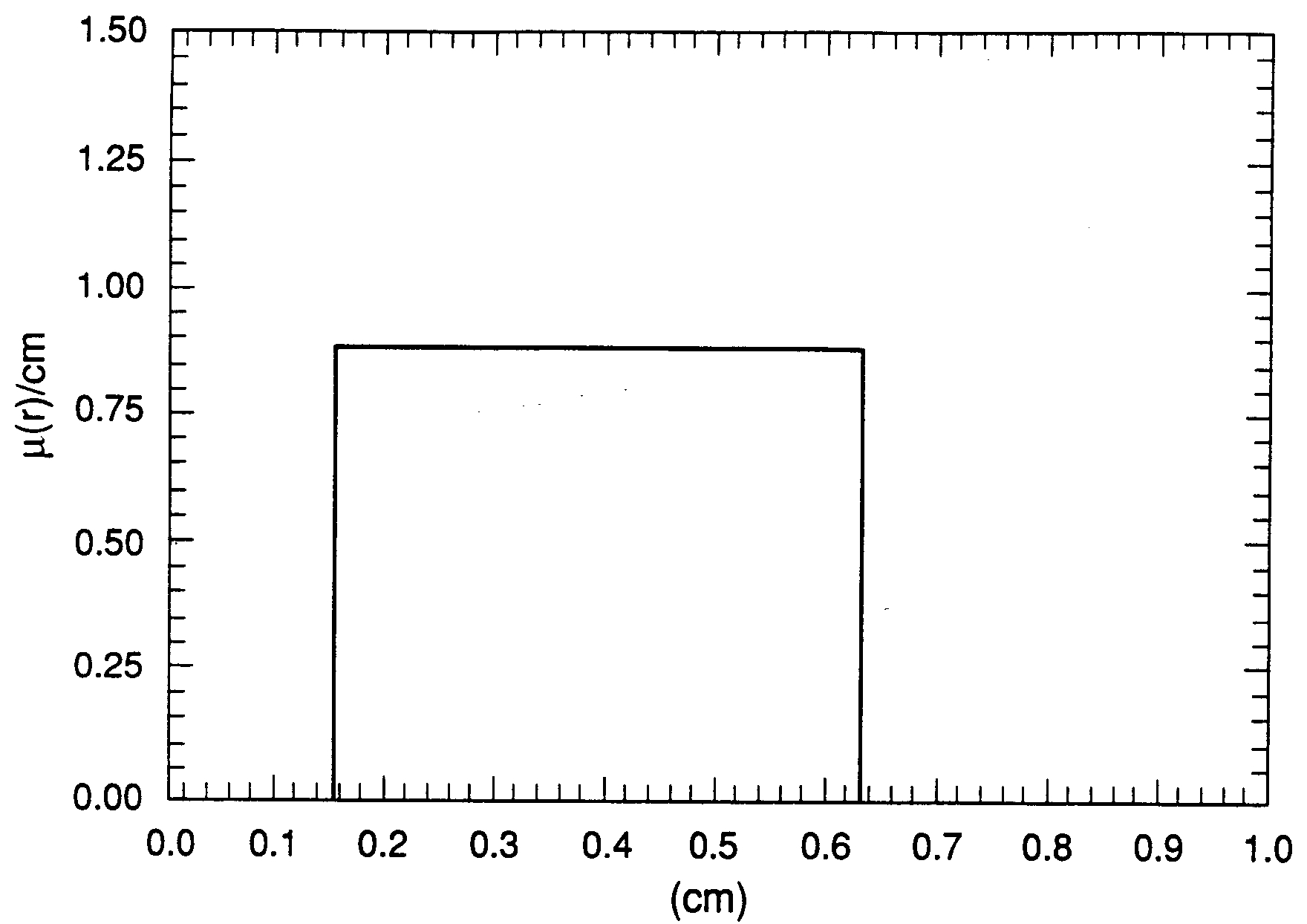} \vspace{0.0truecm}
\caption{The absorption coefficient computed from
Eq.~(\ref{abel2}) for the solid $U^{238}$ using the value given in
Figure \ref{fig:fig6}. The figure shows the code can treat a sharp
boundary with little noise.}
\label{fig:fig10}
\end{figure}

In Figures \ref{fig:fig5} and \ref{fig:fig6}, we show $f(y)$ as a
function of $y$ by Eq.~(\ref{abel}) with $\mu=0.875/cm$ for a
cylinder of $U^{238}$. In order to see the details of discrepancy
between the experimental data and the theoretical values, we have
also computed $f_{exp}(y)$ from the experimental data, which is
displayed in Figures \ref{fig:fig7} and \ref{fig:fig8}.

The quality of $f_{exp}(y)$ should be critically examined by
comparing it with $f(y)$. First we see from Figures
\ref{fig:fig5} and \ref{fig:fig7} that $f_{exp}(y)$ shows a long
tail extending beyond the outer boundary.

Crucial to any x-ray diagnostics is the calibration of the
measured transmitted beam and source intensities. This can be done
either with the theory using a measured absorption coefficient
from a slab geometry or empirically from the measured standards.
The reason for this is the difficulty of estimating the incident
flux intensity $I_{0}$ which varies from shot to shot in a
Bremsstrahlung source. The basic goal in determining $I_{0}$ is to
match as closely as possible the ideal situation of measurements
in the slab geometry.

The long tail in Figure \ref{fig:fig7} clearly suggests that the
calibration for $f_{exp}(y)$ was seriously in error, because
$f_{exp}(y)$ must be zero at the outer surface and become
negative beyond the boundary due to Compton scattering at the
surface.

Next we observe that $f_{exp}$ exhibits the intensity fluctuation
which, we believe, is due to enhanced backscattering in a
disordered medium (Anderson localization \cite{Anderson58}). In
the experimental data on Carbon reported by several different
groups \cite{Creagh87}, one notices a wide range of disagreement
in the absorption coefficients. In an experiment with Carbon, one
must be concerned with x-ray absorption in a disordered medium
({\it i.e.,} due to a random distribution of small voids). We
believe the disagreement can be attributed to Anderson
localization, which implies a breakdown of a model of $N$
independent, identical atoms for a medium. This is a macroscopic
realization of quantum many-body effects in a disordered medium.

Furthermore, the recent experiments of light propagation in a
disordered medium \cite{Wolf85,Alvada85} provide direct evidence
of enhanced backscattering with the light intensity fluctuation
which are very sensitive to the pattern of speckles in the
medium.  These measurements are consistent with our data, which
suggests that the x-ray intensity fluctuations are due mostly to
impurities in $U^{238}$ and partly by the broad spectrum of the
bremsstrahlung source.

If we compare Figures \ref{fig:fig5} and \ref{fig:fig7}, we can
clearly see the effects of progressive filtering and large angle
Compton scattering; the breakdown of the narrow beam
approximation is particularly evident. An important question
concerning the use of x-ray radiography for quantitative
determination of a density profile is how accurately both the
progressive filtering and Compton scattering effects can be
determined and their contribution eliminated. If they can be
accurately accounted for, then, only then, the remaining data can
provide a density profile within the limit of the narrow beam
approximation.

We observe from Figures \ref{fig:fig5} and \ref{fig:fig7} that
the breakdown of the transmission rate, Eq.~(\ref{absorption}),
increases as $r$ approaches the center of a test object. We also
see appreciable effects of Compton scattering at the edge, which
produce a diffuse boundary. This shows that one cannot hope to
isolate these effects by a simple numerical method or background
removal, and thereby obtain precise absorption coefficients by
Abel's inversion. These effects must be corrected by filtering
technique that can keep the spectrum width $\Delta \lambda$ as
small as possible; this eliminates the progressive filtering by a
test object.

By placing an appropriate filter in front of a film, we may
further reduce the effect of Compton scattering at the edge.
Again, what has been said about the Compton scattering effect also
applies to the curved surface of a test object in which an oblique
incidence of an x-ra beam leads to a partial reflection of the
beam. This can be corrected only by an analytic method.

Comparing Figures \ref{fig:fig6} and \ref{fig:fig8} we observe
similar effects in $f_{exp}(y)$ for a hollow cylinder, but a large
discrepancy between $f_{exp}(y)$ and $f(y)$  exists near the
center extending to the boundary of the void. In the absence of
dense material at the center of the test object, the peculiarly
strong absorption of x-rays at the void can be explained only in
terms of the multiple reflection of soft x-rays at the the sharp
boundary ({\i.e.,} a radiation trap), as will be discussed later.

The absorption coefficient $\mu$ in the numerical calculation of
$f(y)$ was measured in the slab geometry in which the surface
scattering effects are subtracted by adjusting
$I_{true}=I_{0}-\delta I_{0}$ where $\delta I_{0}$ is a correction
term. Although considerable accuracy of the experimental values
has been achieved from such measurements, similar measurements
using an object with a geometrical structure do not provide the
same degree of accuracy, partly because we are not able to find a
reasonably accurate analytical form for $\delta I_{0}$.

Contrary to common belief, the significant discrepancy between
$f_{exp}(y)$ and $f(y)$  in the presence of a void in materials
does not involve spatial resolution in connection with the spot
size of an x-ray source, but is instead due to both Compton
scattering and the progressive filtering. This is evident because
we can still locate the boundaries of a test object whose
dimension is less than the spot size.

Finally we have calculated the absorption coefficients for both a
sold cylinder Figure \ref{fig:fig9} and a cylinder with a void at
the center  Figure \ref{fig:fig10} using the numerical code
developed based on the Eqs.~(\ref{last}). The results clearly show
that Eqs.~(\ref{last}) are indeed a correct solution to the Abel's
integral equation Eq.~(\ref{abel}) and the numerical accuracy of
the code is indeed impressive.

\subsection{\label{sec:level2} Computation of Absorption
Coefficients from Experimental Data}

\begin{figure}[tbp]
\includegraphics[bb=0in 0in 6in 4in,totalheight=4in,height=4in,width=6in,
keepaspectratio=true]{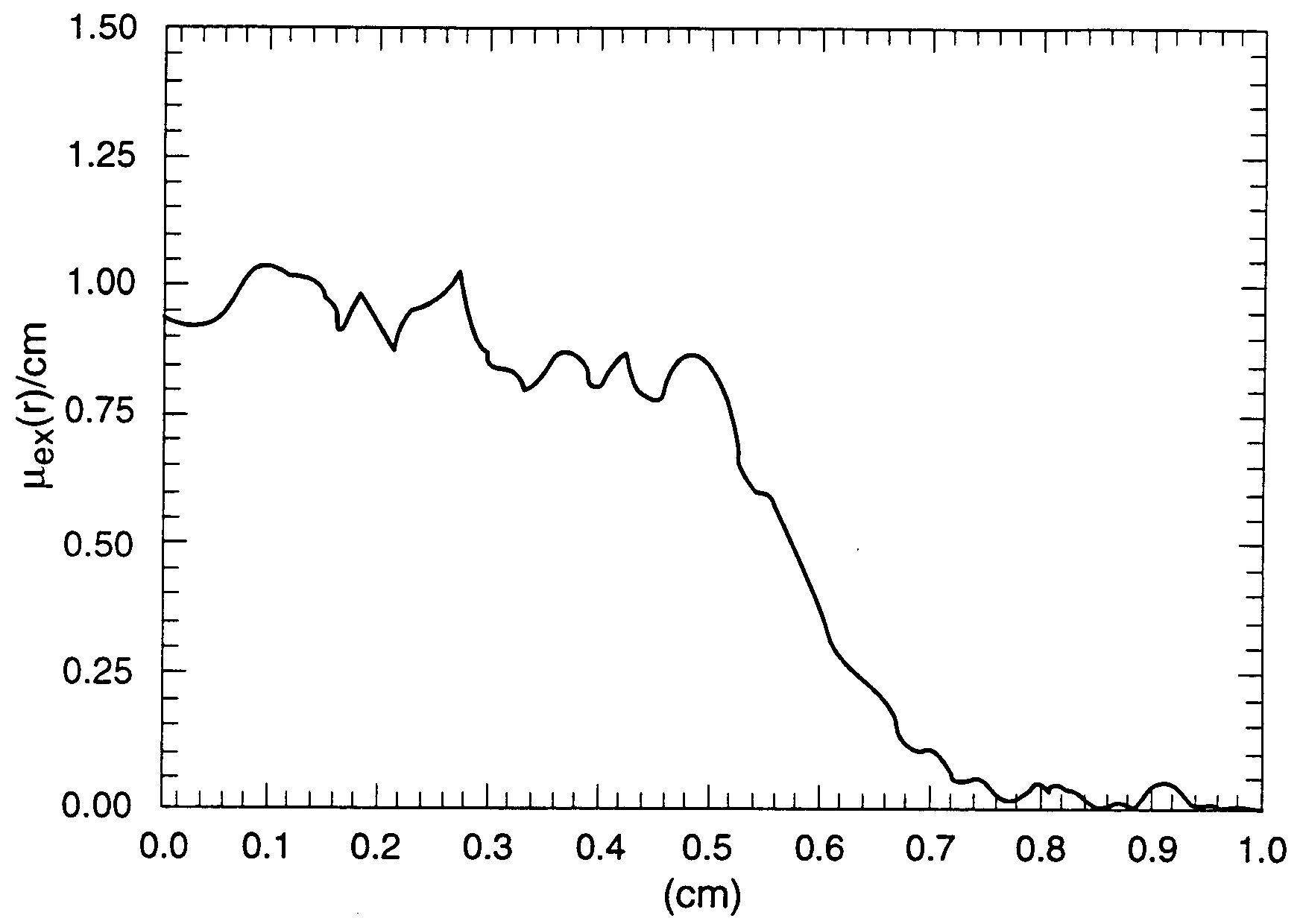} \vspace{0.0truecm}
\caption{The absorption coefficient computed from the experimental
data on the $U^{238}$ solid cylinder. This figure shows the x-ray
intensity fluctuations due to impurities in $U^{238}$ and a
diffuse boundary. The lobes represent the interference effects of
scattered x-rays with the spatially uniform incident x-rays.}
 \label{fig:fig11}
\end{figure}

\begin{figure}[tbp]
\includegraphics[bb=0in 0in 6in 4in,totalheight=4in,height=4in,width=6in,
keepaspectratio=true]{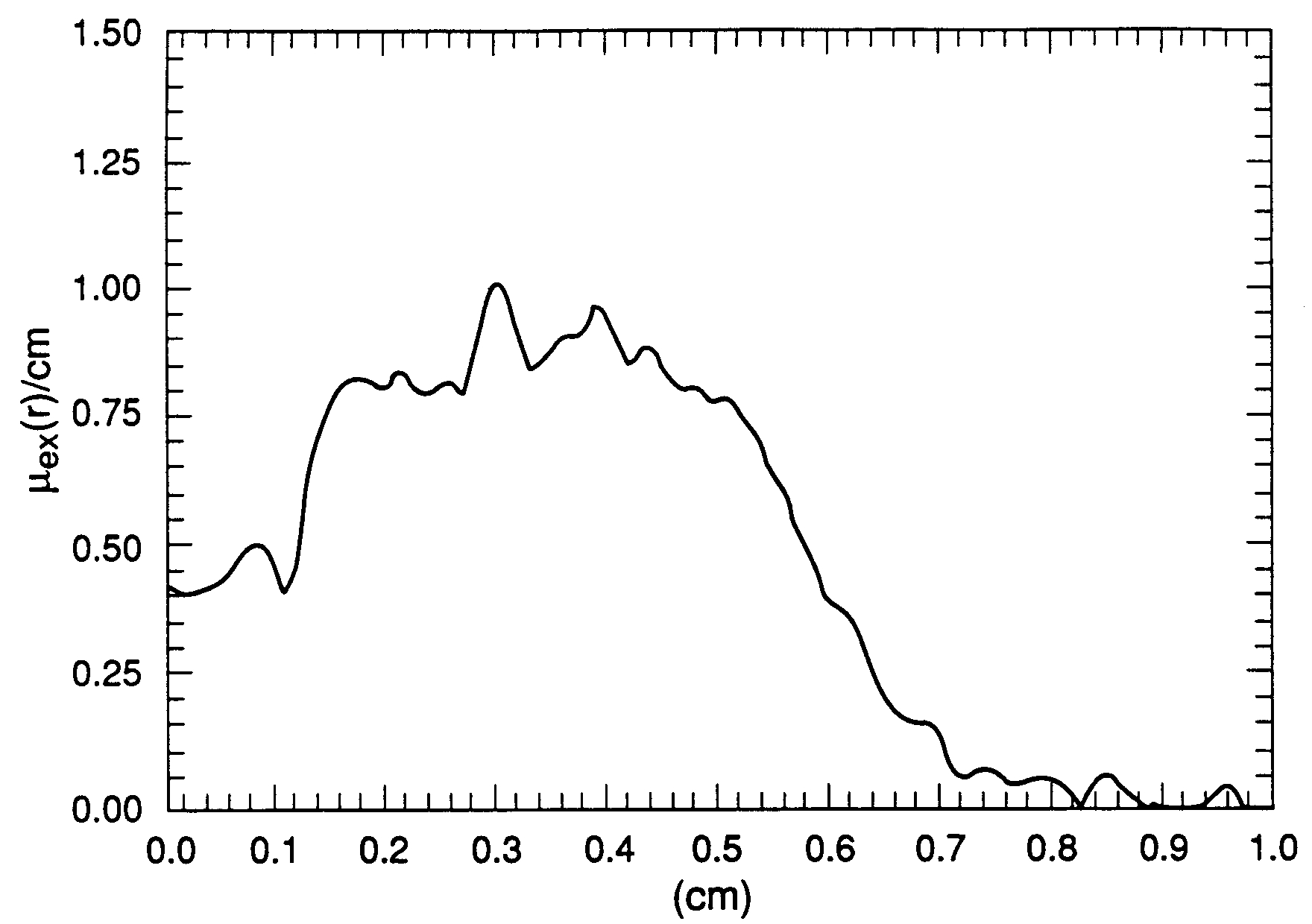} \vspace{0.0truecm}
\caption{The absorption coefficient computed from the experimental
data on the cylindrical shell of $U^{238}$. The large absorption
coefficient $\mu(r)$ near the center shows an excessive x-ray
absorption due to the presence of a void. } \label{fig:fig12}
\end{figure}

By means of Abel's inversion Eqs.~(\ref{last}), the absorption
coefficients are computed from the experimental data with the code
and are shown in Figures \ref{fig:fig11} and \ref{fig:fig12}.
Analyzing the data from the computed absorption coefficients has
the advantage by comparing with Figures \ref{fig:fig9} and
\ref{fig:fig10} in that one clearly sees the effects of enhanced
backscattering and Compton scattering.

The main result obtained from this numerical computation of the
absorption coefficients is the demonstration of a strikingly
accurate numerical algorithm that treats the random noise without
any sign of amplification. Moreover, the calculated absorption
coefficients show a consistency with the experimental data.

Figure \ref{fig:fig11} shows the calculated absorption coefficient
as a function of radius for the solid cylinder of $U^{238}$.
Several important qualitative conclusions can be drawn from the
figure. Fist of all, the variation of the calculated absorption
coefficient exhibits irregular patterns containing a smooth
oscillation in each isolated region, a phenomenon that can only
arise as a result of enhanced backscattering in a disordered
medium. Moreover, the calculated absorption coefficient reflects a
small but pronounced local fluctuation in the x-ray intensity in
the medium.

As mentioned earlier, such an effect has also been observed in
the experiments on the multiple scattering of visible light by an
aqueous suspension of sub-micrometer size mono-disperse
polystyrene  spheres \cite{Alvada85,Wolf85}. These experiments
confirmed the enhanced backscattering of light waves by
disordered scattering centers in a medium.

A qualitative picture of the dependence of the correction term
$\delta I_{0}$ on the glancing angle can be inferred, for the
energy region of our interest, from the surface reflection
phenomenon at x-ray energies less than 10~keV \cite{Parratt54}.
The shape of $\delta I_{0}$ sensitively depends on the incident
photon energy and glancing angles less than the critical angle
(above which there is no reflection). By adjusting the correction
term to give the theoretical values $f(y)$ in closer agreement
with the experimental data, one learns which physical processes
are responsible for the various peculiarities in the data.

For the same cylinder with a void, the absorption coefficient
calculated from the experimental data is shown in Figure
\ref{fig:fig12}. The effect of a partial reflection of x-rays at
the boundary of a void can be clearly seen. The error in the
absorption coefficient is about 52 \% at the center, which appears
to suggest the presence of radiation absorbing material. This
significant disagreement between the computed value and the
expected value (Figure \ref{fig:fig12}) reflects the neglect, in
this simple form of absorption mechanism, of the fundamental
physics treating a void in a medium or suggests an error in the
calibration of the measured transmitted beam and source
intensities, or both. As explained above, the calibration of the
experimental data is clearly in serious error. A proper
calibration would have reduced the error in the absorption
coefficient to less than 30 \%.

Any simple physical process that may have been involved in this
x-ray transmission is not abstruse, since one notices that the
reflection of an incident x-ray beam at the outer and inner
surfaces can reduce the transmitted beam intensity significantly.
The maximum reflection takes place at the edge of a void where
the glancing angle becomes nearly equal to the critical glancing
angle \cite{Parratt54}. As the glancing angle increases, the
effect of x-ray reflections reduces. It should be emphasized that
a bremsstrahlung source is far from being monochromatic as we
will discuss later. Thus soft x-rays, whose critical glancing
angles are much larger than those of hard x-rays, will reflect
more readily at the surfaces. For high-energy photons, the
angular dependence of scattered photons must be analyzed by the
Klein-Nishina formula \cite{Bjorken64}.

It is straightforward to assess the effect of Compton scattering
at a sharp boundary for a monochromatic source, but it becomes
prohibitively difficult for a bremsstrahlung source. Thus we can
see that Compton scattering at a sharp boundary leads the
breakdown of the basic assumptions of x-ray transmission in the
void for which EGS4 can be utilized to see the effect of the
bremsstrahlung source. In passing we note that, in Compton
scattering experiment, a target of circular cross section such as
cylinder is often used for the study of angular dependence; the
scattering center at the surface is better defined even if the
beam is poorly collimated. However, as we have seen in our data
analysis, accurate evaluation of the flux intensity at the target
is difficult.

We are thus faced with a question: how does the x-ray reflection
at the surface of a void differ from that of the outer surface?
Moreover, in the presence of photo-emitted electrons, how are we
to describe the x-ray transmission in the void? To answer these
questions we must focus on the nature of interactions of photons
at the surface and in the free space of the void.

The incident x-rays see the concave surface, where a partial
reflection of the primary and secondary photons can take place. A
fraction of the primary and secondary photons are ultimately
absorbed near the boundary of a void, giving rise to enhanced
photoemission of electrons in the void. We also note that, in the
absence of a strong Coulomb field of nucleus, no pair-production
can occur in the void.

Furthermore, Compton scattering of primary and reflected photons
by photo-emitted electrons in the void is small compared with that
of condensed matter, but it can be a contributing factor to the
discrepancy, since the trapped photon intensity is much larger
than the primary x-ray beam intensity for a small void.

In general, as expected, the effect of the x-ray reflections on
the transmitted beam intensity is less severe as the radius of a
void increases. This can be easily understood by the fact that the
x-ray reflection takes place near the edge. An important point to
notice is that near the edge of the void we see a significant dip
in the absorption coefficient which shows the reflection of
x-rays. In summary, {\it we have shown that a numerical
calculation of density profile by Abel's inversion has its
limitation and cannot be used as a quantitative tool for a
hydrodynamic test!} The conventional approach to examine the film
by experienced eyes will provide a better diagnosis for a
hydrodynamic test. The noise is inherently too large in its
magnitude to neglect. In addition, a three-dimensional density
profile of a test object cannot be obtained in an environment of a
dynamical test of hydrodynamics as discussed below.

\section{\label{sec:level1}Optimal Experimental Conditions}

In order to obtain reasonably accurate absorption coefficients
from the experimental data  by means of Abel's inversion, the
most optimal experimental condition must be realized so that the
data to be analyzed are free of significant distortion.

To derive the Abel's integral equation Eq.~(\ref{abel}), several
simplifying assumptions are made. We examine the assumptions to
minimize the errors that they introduce and to find the optimal
experimental conditions.

The most serious approximation made was to assume that a
bremsstrahlung source is monochromatic. The potential difficulty
introduced by this assumption can be appreciated if we recall
that a bremsstrahlung source spans the entire spectrum exhibiting
a relatively flat intensity profile with respect to photon energy
\cite{Motz59}. Furthermore an absorption coefficient for any
given material depends on photon energy and Z-number. Thus it is
important to limit the width of the spectrum to the smallest
possible value.

There are several technique for limiting the wavelength range
$\delta \lambda$ available, with each its advantages and
disadvantages. The simplest method is to use a crystal
monochromater. However this would be impractical in our x-ray
diagnostics, because the  wavelength of our x-rays is too short
compared with a lattice parameter of available material.

From Fig. 32 of Bethe and Ashkin \cite{Bethe53} we see that there
is a region of maximum transparency for all material (MTM) which
lies between 2.5~MeV and 10~MeV. This is admittedly a rather
loosely defined spectral range, its limit being in principle
different for each metal, but one may think of the spectral range
as the window of maximum transparency for heavy materials. In the
language of scattering theory, one can say that while at high
energies ({\it i.e.,} ${\cal E}_{\gamma} \gg 10 MeV$) the mean
free path is so small that for a thick target a photon must
undergo multiple scattering to reach a detector. As the photon
energy is lowered into the region of MTM the photon mean path is
sufficiently large that the effect of multiple scattering can be
negligible. As discussed earlier, the transmission rate defined in
Eq.~(\ref{absorption}) does not admit multiple scattering, which
can be a serious effect for a thick targets. Because the x-rays in
the region of MTM can penetrate to a great depth without
significant multiple scattering, a varying photon energy $\langle
{\cal E}\rangle_{av}$ can cover many blemishes in Abel's integral
equation. Thus the solution of the integral equation can give a
reasonably reliable absorption coefficient apart from the large
angle Compton scattering which can be eliminated without
difficulty.

This discussion has an important implication with respect to a
design of an x-ray source for an x-ray diagnostics. The basic
goal in selecting $\langle {\cal E}\rangle_{av}$ is to match as
closely as possible the upper limit of the region of MTM with the
smallest possible spectral width. Since the maximum photon energy
in a bremsstrahlung source is nearly equal to an incident
electron beam energy hitting a tungsten target, the upper limit
can be set by selecting a specific electron energy.

A better technique than a monochromater for reducing the spectral
width is to use filtering materials by which the photons with
energy less than 5~MeV are eliminated and Compton effects are
minimized. It is important to note that since the same absorption
coefficient can be measured at two different energies in this
region, we must block the low energy photons before they reach a
test object. This is possible because the x-rays to be filtered
are of higher energy than the fluorescent signal.

By choosing an appropriate filter, its absorption edge can be
placed between the photon energy (to be filtered) and fluorescent
energies, preferentially absorbing the former. The major
difficulty with this method of x-ray filtering is that absorbed
x-rays can be re-emitted as fluorescent radiation that is
relatively un-attenuated by filters and strike film to increase
the fog level. This degrades the sensitivity to low level
signals. It should be emphasized, however, that by eliminating
the soft photons we reduce both progressive filtering and Compton
effects at the surface of a test object and that the filtering
improves the quality of data considerably.

On the other hand, the filtering reduces the source brightness,
which is defined as the number of photons emitted per unit solid
angle. Empirically it is found that the bremsstrahlung radiation
intensity in terms of the electron beam energy in the forward
direction is given $I_{\gamma}\propto I_{e}{\cal E}^{2.9}_{kin}$,
where $I_{e}$ is the electron beam intensity and ${\cal E}_{kin}$
is the electron kinetic energy in unit of $mc^{2}$ \cite{Motz59}.
Here the target material was gold. We can appreciate from the
above empirical formula and Fig. 30 of Koch and Motz
\cite{Motz59} that, for example, at a given electron beam energy
it is necessary to increase the current intensity in order to
maintain the same effective source brightness.

This requirement poses a difficult problem for which the
accelerator community is in search of an answer - a new
acceleration mechanism capable of generating an intense
relativistic beam. Fortunately, a novel acceleration developed by
Friedman \cite{Friedman74, Friedman89} has been tested in
experiments demonstrating that a beam of electron current of 70~
kA can be accelerated to a energy of 7~MeV. However, it was found
that the acceleration mechanism is not scalable to higher energy
due to various instabilities.

The unique feature of the auto-acceleration mechanism is that the
electromagnetic field for the acceleration is created by the
interaction of the first electron beam with a cavity. The
electromagnetic field energy is then transferred to the second
electron beam in the same structure, producing a short pulsed
({\it e.g.,} 20 ns), intense electron beam. Moreover, the
auto-accelerator circumvents the two major problems in pulsed
power accelerators - the fast switches and insulators; the
accelerator is automatically self-synchronized and thus
eliminates the jitter due to the switches.

It should be stressed, however, that the efficiency of the
auto-acceleration mechanism depends on the ratio of the first beam
intensity to that of the second. Since the extracted
electromagnetic field energy from the first beam is the only
source of the acceleration for the second beam, the larger the
ratio, therefore, the more efficient the auto-acceleration is
bound to become.

Consideration of the stability of an intense relativistic
electron beam in a structure led Friedman \cite{Friedman89} to
the modification of previous experiments. In a recent experiment
\cite{Friedman89}, the first beam was modulated by using an
external rf source to enhance the interaction between the beam
and the modified structure. Hence the accelerated beam and the
modulated beam exchange energy through the modified structure
while moving on different trajectories. Friedman has demonstrated
that the accelerated beam of 200~\AA~ attains the kinetic energy
of 60 MeV with the beam radius of 0.1 cm. This result strongly
suggest that there exists a domain of parameters in which we can
select an appropriate accelerated beam intensity for an optimally
useful x-ray source.

We now examine in some detail the characteristics of of an intense
electron beam for an x-ray source for the radiography: first, we
study the efficiency of bremsstrahlung production for a given beam
energy, which is defined as the ratio of the total bremsstrahlung
radiation emitted to the total energy of an incident electron beam
and is given by $\epsilon=3.0\times10^{-4}Z{\cal
E}_{0}/(1.0+3.0\times 10^{-4}Z{\cal E}_{0})$ where ${\cal E}_{0}$
is the kinetic energy of an incident electron \cite{Motz59}. The
efficiency depends on the target material and the kinetic energy
of an incident electron beam. For example, the efficiency
$\epsilon=31$ \% at ${\cal E}_{0}=20 mc^{2}$ but it is 41\% at
${\cal E}_{0}=30 mc^{2}$ for Pt targets. Note also that the
spectrum shape of bremsstrahlung intensity is given by
$S=1-\exp{-{\cal E}_{0}/[51({\cal
E}_{0}-\epsilon_{0})\epsilon^{2}_{0}]}$, where ${\cal
E}_{0}-\epsilon_{0}=h\nu$, $\nu$ is the frequency of a photon
emitted in the forward direction, and $h$ is Plank's constant
\cite{Motz59}.

Thus for 10~MeV electron beam incident on a thick tungsten target,
the efficiency $\epsilon=30~\%$ and about 35~\% of photons emitted
in the bremsstrahlung process have their energy higher than 5~MeV.
In the x-ray transmission, unfiltered soft photons are absorbed
through progressive filtering in a test object and do not reach a
detector. Assuming the spectrum width $\delta\lambda=1.24\times
10^{2} \AA$, which is equivalent to 1~MeV, the optimally useful
x-rays are, therefore, 15~\% to 25~\% of the total photons emitted
in the bremsstrahlung processes.

At extremely high electron beam energies like the one in the
PHERMEX facility (the electron beam current ${\cal I}_{e}=2.8 kA$
and the electron energy (${\cal E}_{0}=30 MeV$), the radiation
cone from a bremsstrahlung source is extremely narrow and x-ray
photons are emitted mostly in the forward direction. Thus spectrum
shape can be described  by the above analytic expression which
shows the peak intensity in the energy range of 5~MeV to 6~MeV. It
should be remarked, however, that although the efficiency for this
beam is $\epsilon=57$~\%, hard $\gamma$-rays from a bremsstrahlung
process tend to cascade in a test object and the x-ray
transmission is controlled mainly by the photons with energy in
the neighborhood of the peak intensity and hardest secondary
photons are produced in multiple scattering. These examples are
given to show certain features that provide a qualitative picture
to identify an optimally useful x-ray source. Again, one may try
to use EGS4 code to check if the above qualitative picture holds
for a target design of a bremsstrahlung source.

The above analysis shows that the high-current, high-energy
electron beam is obviously preferred for an optimally useful
x-ray source and the only remaining question that concerns if the
photon energy falls in the region of MTM.

The second issue is the more fundamental problem, which is common
in all high-current accelerators; it is important to focus an
electron beam to obtain uniformly, diverging x-ray beam to
satisfy the assumption of a parallel x-ray beam. Although it is
extremely difficult, one should be able to achieve a small spot
size by the combined use of a quadrupole magnet and ion focusing
\cite{Nezlin68,Han1}.

We also note that the shape of radiation cone from bremsstrahlung
processes depends on the incident electron beam energy and that,
with an electron beam energy in the region of MTM, the radiation
cone is sufficiently large that the x-ray beam diverges uniformly
in space. This is a critical factor for the spatial uniformity of
the x-ray beam, which is the basic assumption in deriving Abel's
integral equation.

In the light of the above problem, it is essential to develop a
simple, reliable acceleration mechanism such as Friedman's
auto-acceleration mechanism. Indeed, the simplicity and the
efficiency of the auto-acceleration mechanism hold out the hope
that one might develop an x-ray source that meets the
requirements. Clearly the auto-acceleration mechanism shows
definite advantages over a more conventional rf linac for
generating a high-current, high-energy electron beam. Moreover,
this type of accelerator is compact and inexpensive to build
compared to an rf linac.

We would like to emphasize, at the end of this detailed
discussion on the optimal condition for an x-ray source, that the
physics and experimental technique elaborated here provide a
starting point from which we can develop an optimally useful
x-ray source with narrow spectrum width and adequate beam
intensity.

\section{\label{sec:level1}Discussion and Conclusion}

Although our analysis of the experimental data is qualitative in
nature, we have attempted to quantify the discrepancy from the
form of transmission rate $I=I_{0}e^{-\mu x}$, by using Abel's
inversion Eqs.~(\ref{last}) and its numerical solution. Given the
large uncertainties due to the lack of detailed understanding of
the physical processes in the x-ray transmission for a given test
object with a structure as well as the limited data, our
interpretation should be taken as only suggestive.

One of the major problems that confronted us in this experiment
was the complexity of the acquisition and interpretation of the
experimental data. Abel's inversion is conceptually simple and
mathematically straightforward. Qualitative though this picture
may be, as it has to be, since a simplified transmission of
$\gamma$-rays in material cannot hold true even in its first-order
analysis. The lack of detailed agreement between the theory based
on the a simple x-ray transmission rate and the experimental data,
however, shows a complexity of transmission of radiation in high
Z-material. Since the x-ray transmission itself is inherently
complex a phenomenon, there seems little hope to achieve the
reliability required for a quantitative method in x-ray
diagnostics in hydrodynamics. {\it Thus it is highly unlikely that
the quantitative diagnostic by the radiographic method in
hydrodynamic test can be meaningful.} In practice, it can be
misleading.

That the experimental measurements are not always reliable
themselves is not an important issue, since the most optimal
condition can be realized, and perhaps reliable data can be
obtained. Even under the best condition, the data cannot still
satisfy the basic assumption on Abel's inversion. In general,
several different corrections must be made to the directly
observed transmitted x-ray intensity in order to obtain a
reasonably credible absorption coefficient by Abel's inversion.

It should be emphasized, however, that, in most experiments,
$I_{0}$ and $I(x)$ are read simultaneously from a film, and that
the incident and the transmitted x-ray beam intensities are
assumed to be nearly monochromatic, which is far from being true.
This requirement of an x-ray source is essential for the
measurement of absorption coefficients, which depend on the
average energy of an incident photons. Numerous simplifying
approximations ({\it e.g.,} the narrow beam approximation, {\it
etc.}) have been made for reasons of both availability of an x-ray
source and expediency. There is no practical method to produce
monochromatic x-ray beam. The final effects of these
approximations remain unknown.

Application of Abel's inversion in x-ray diagnostics is not as
straightforward as for plasma diagnostics since we do not know
how to make corrections to the transmitted x-ray intensity.
Sometimes one might get the impression that the correction to the
transmitted beam intensity, $\delta I$, is something of an art.
For the worst case this is probably true.

One must therefore exercise the scientific prudence by checking
every step with the Monte Carlo code EGS4 \cite{EGS4} to insure
that the corrections are well justified and are in the order of
procedures weighted by the importance of the processes: a) an
incident x-ray beam intensity; b) removal of background due to
photon cascade; c) Compton scattering at a sharp boundary; d)
progressive filtering due to a finite spectrum width; e) a narrow
beam approximation.

The relative importance of each of these corrections, as well as
the details of correction procedure, depends on the x-ray beam
energy and the particular material composition of a test object.
Unfortunately, little discussion of these corrections has been
attempted in most of the published measurements of x-ray
transmission. This has greatly impeded progress in x-ray
diagnostics.

A crucial question concerning the application of Abel's inversion
for quantitative x-ray radiography is how accurately the Compton
effects such as partial reflection of x-rays at a sharp boundary
and x-ray scattering at the edge can be determined and can be
corrected. Unfortunately the most difficult quantity to measure in
x-ray radiography is an accurate incident x-ray beam intensity
$I_{0}$. A variety of effects can distort the incident beam
intensity profile severely while leaving the energy spectrum
relatively undisturbed. Yet it is the incident beam intensity that
is critical factor in determining the absorption coefficients by
Abel's inversion.

As explained earlier, the reflected x-ray beam intensity at a
sharp boundary varies with the incident glancing angles larger
than the critical angle $\phi_{crt}$. This implies that $\delta
I$ with a test object having a curved surface depends on the
geometrical structure factor which is extremely difficult to
determine.

Any background removal method has the potential of distorting the
data. It is therefore important to isolate each contribution to
the correction term by combined use of analytic method and EGS4
code. For example, some of the strongest Compton scattering
effects occur at the outer edge, where they are caused by the
total reflection of an x-ray beam. Occasionally the simplest
remedy for this Compton effect is, albeit difficult in some
instances, to place an appropriate filter in front of a film. One
can see the effect with the clarity of an improved image.

On the other hand, with the data having contributions from
progressive filtering of photons, it may be difficult to find a
simple remedy that can isolate the filtering effects. One notices
from the absorption curves of metals (see Fig. 32 of Ref.
\cite{Bethe53}) that one should not use an average photon energy
for the calculation of absorption coefficients unless the
spectrum width of an x-ray source is small. One should have
expected that, in addition to the progressive filtering, the
narrow beam approximation [Eq.~(\ref{absorption})] will break
down in this most unfavorable case, {\it i.e.,} a bremsstrahlung
source without filtering.

The difference between the absorption coefficients calculated
from experimental data taken for a cylindrical test object and
the experimental value determined in slab geometry is quite
significant, suggesting that the major effect ({\it e.g.,}
Compton scattering by free and bound electrons at large
scattering angles) in the presence of a sharp boundary was not
taken into account.

Because of this, it is essential to find the method that is
consistent and systematic in the selection of parameters for
$\delta I$. This can be achieved by analytic study of the
reflection and absorption of x-rays at the surface of a test
object and systematic experimental tests for various test
objects. It is imperative, albeit tedious, to make several
attempts at background removal with different sets of parameters,
follow each set of experimental data through the entire analysis
for different test objects, and test the results for
consistency.

Although the calculation of $f_{ex}(y)$ is an intermediate step
in Abel's inversion, a great deal of information may be gained by
direct inspection. For test objects with small amount of
impurities, the theoretical value $f(y)$ calculated from
Eq.~(\ref{abs2}) using the experimental value $\mu$ determined in
slab geometry can be used for a comparison with $f_{ex}(y)$. This
comparison will show the fluctuations in the transmitted x-ray
intensity and diffuse boundaries. This exercise is useful as an
overview of the data, yielding the information about a possible
form of correction to the incident x-ray intensity. Inspection of
$f_{ex}(y)$ is also useful as a check of the x-ray beam profile.

The fluctuations in x-ray beam intensity, which should not be
confused with statistical fluctuation of noise \cite{Rose48}, are
intrinsic phenomena which one should not apply a smoothing
technique in any data analysis; the fluctuations reflect the
property of a material, for example Anderson's localization by
impurities.

Next, it should be stressed that Abel's integral equation derived
from the simple x-ray transmission rate Eq.~(\ref{absorption})
does not admit multiple scattering which can modify the
transmission equation significantly. As a general guide in the
design of experimental x-ray transmission measurements, we may
find it useful to check if the dimension of a test object is less
than $5\lambda_{mfp}$, the mean free path of a photon at a given
energy. For a thin target, the multiple scattering can be safely
neglected. One may not able to neglect the multiple scattering,
however, in a thick target  ({\it e.g.,} a size larger than $10
\lambda_{mfp}$). The error introduced in Eq.~(\ref{absorption})
by the multiple scattering will enter into the later calculation
of the absorption coefficients of materials by Abel's inversion.
Such a check can be easily made by referring to a published x-ray
absorption coefficient table for metals and a systematic test
with EGS4 code.

Finally, we have shown how we may obtain one-dimensional density
profile by computing the absorption coefficient with Abel's
inversion Eq.~(\ref{last}) from two-dimensional x-ray film for a
cylindrically symmetric object in one x-ray exposure for which an
x-ray source is located in a sufficiently far distance from a test
object to insure a parallel beam. There is little need to see a
two-dimensional density profile in a radiography, since we can
easily obtain the two-dimensional view in one x-ray exposure. A
question naturally arises, however, how one may obtain a
three-dimensional density profile that is quite useful in
non-invasive diagnosis such as in medicine. Since the wave number
$k_{\theta}=m/r$ in cylindrical coordinates, the spatial
resolution requires that {\it the $m$ exposure by rotating the
x-ray source about the axis ($z$-direction) to obtain the
three-dimensional density profile.} This requirement was
misunderstood in his study of a line integral for application of
to radiological application by Cormack \cite{Cormack63} and thus
his analysis is invalid. The $m$ different views provide the
necessary information on density profile in each direction which
can be used to construct the three-dimensional density profile as
in a three-dimensional radiographic machine. For a given mode
number, the best spatial resolution will be achieved near the axis
of symmetry. A rule of thumb on the spatial resolutions is: the
larger the number of exposures $m$ is, the more reliable the
three-dimensional density profile will be in a numerical
reconstruction \cite{Hamming62}. This can be achieved either by
rotating an x-ray source or rotating a test object, which is not
feasible in any dynamic hydrodynamic test. This analysis does
clear up the myth that one may be able to build a 3-D hydrodynamic
test facility with the two x-ray sources which is known as the
Dual Axis Radiographic Hydrodynamic Test Facility or DARHT project
at Los Alamos \cite{comment3}.

\newpage

\begin{thebibliography}{95}
\bibitem{Comments1} One should be aware of the assumption of a
        paralell incident x-ray beam which is produced at a far
        distant by a point source. Hence one cannot apply a Fourier
        decomposition on the angular dependence in Cormack's Equation 2.
\bibitem{Comments2} This optimal condition does not allow the Fourier
        decomposition of Abel's integral equation as was carried
        out by Cormack. Hence Comack's analysis is not valid for
        an optimal condition of radigraphy.
\bibitem{comment3} This assumption is critical to the derivation
        of Abel's inversion (CAT scan). Hence the three
        dimensional CAT scan is feasible only if a test object is
        rotated with sufficiently small angular resolution or the
        x-ray source is rotated about the axis of symmetry. One
        may apply the Fourier decomposition with N-different
        views, since each view provides a one dimensional density
        profile as shown below.
\end{thebibliography}
\end{document}